\documentclass[floatfix,superscriptaddress,aps,prb,showpacs,twoside,twocolumn,10pt]{revtex4-2}
\usepackage[colorlinks=true, citecolor=blue, urlcolor=black ]{hyperref}
\usepackage{epsfig,newlfont,amssymb,amsfonts,amsmath,bm,subfigure,palatino,mathtools,amsthm,braket,times,soul,enumitem,color}
\usepackage[normalem]{ulem}
\usepackage{xcolor}
\usepackage{cancel}
\usepackage[T1]{fontenc}
\usepackage[normalem]{ulem}
\usepackage{graphicx}
\usepackage{placeins}
\usepackage{hyperref}
\usepackage{tikz}
\usepackage{multirow}
\usepackage{hhline,booktabs}
\usepackage{textcomp}

\usepackage{blindtext}
\usepackage{graphicx}
\usepackage{amsmath}
\usepackage{bm}
\usepackage{mathrsfs} 
\usepackage{hyperref}
\usepackage{geometry}
\usepackage{amsthm}

\usepackage{gensymb}
\usepackage{physics}
\newcommand{\ab}{\color{black}}

\newcommand{\gb}{\color{black}}

\begin{document}

\title{{\gb{Reduced State Stabilizer R\'enyi Entropy}} as a Probe of Quantum Phase Transitions in Frustrated \(J_1\!-\!J_2\) Spin Models}



\author{George Biswas}
\email{georgebsws@gmail.com} 
\affiliation{Department of Physics, Tamkang University, Tamsui Dist., New Taipei 25137, Taiwan, ROC}

\author{Santanu Sarkar}
\affiliation{Department of Physics, National Institute of Technology Sikkim, Ravangla, Namchi, Sikkim 737139, India}

\author{Jun-Yi Wu}
\affiliation{Department of Physics, Tamkang University, Tamsui Dist., New Taipei 25137, Taiwan, ROC}
\affiliation{Hon Hai Research Institute, Taipei, Taiwan, ROC}
\affiliation{Physics Division, National Center for Theoretical Sciences, Taipei, Taiwan, ROC}

\author{Anindya Biswas}
\email{anindya@nitsikkim.ac.in} 
\affiliation{Department of Physics, National Institute of Technology Sikkim, Ravangla, Namchi, Sikkim 737139, India}

\date{\today}

\begin{abstract}
We investigate whether the second-order purity-corrected stabilizer R\'enyi entropy (SRE) of reduced two-qubit density matrices can serve as a reliable local indicator of quantum phase transitions (QPTs) in frustrated quantum spin systems. We consider the one-dimensional isotropic \(J_1\!-\!J_2\) Heisenberg model, the one-dimensional XXZ \(J_1\!-\!J_2\) model, and the two-dimensional \(J_1\!-\!J_2\) Heisenberg model on a \(4\times4\) square lattice. Unlike several previously studied quantum information measures, which fail to detect QPTs in the ground state of these frustrated systems, the reduced ground-state purity-corrected SRE successfully identifies most transitions. For the remaining cases, we consider a low temperature subjacent state, modeled as a statistical mixture of the ground and first excited states with a Maxwell--Boltzmann-type occupation probability. For the 1D isotropic model, the subjacent-state SRE shows a discontinuity at the critical point, yielding \(\alpha_c(\infty)=0.24116\), in excellent agreement with established values; the ground-state SRE shows a point of inflection, yielding \(\alpha_c(\infty)=0.2681\). For the 1D XXZ model, the subjacent-state SRE reproduces the full anisotropy dependent phase diagram, while the ground-state SRE captures transitions only at low anisotropy. For the 2D model, the subjacent-state SRE detects two transitions, at \(\alpha_c(4\times4)=0.40781\) and \(0.6208\), while the ground-state SRE identifies the second at \(0.6230\). Compared with conventional two-qubit entanglement, purity-corrected SRE shows a clear advantage in revealing otherwise-invisible phase transitions, establishing it as a robust, efficient, local probe of frustrated quantum criticality.
\end{abstract}

\maketitle

\section{Introduction}\label{intro}

The study of quantum many-body systems through the lens of quantum information theory has emerged as a powerful approach for understanding collective quantum phenomena, especially quantum phase transitions (QPTs)~\cite{sachdev_2011}. 
Quantum phase transitions occur at absolute zero temperature when a qualitative change in the ground-state properties of a many-body system is induced by the variation of an external control parameter or interaction strength~\cite{sachdev_2011,Carr2011,Dutta2010}. Unlike classical phase transitions, which are driven by thermal fluctuations, quantum phase transitions arise from quantum fluctuations and changes in many-body correlations. This intimate connection between quantum information and many-body physics has motivated the development of numerous information-theoretic probes of criticality~\cite{PhysRevA.66.032110,Osterloh2002,Gu_2006,PhysRevLett.90.227902,PhysRevA.70.042311,PhysRevLett.93.250404,dutta_aeppli_chakrabarti_divakaran_rosenbaum_sen_2015,PhysRevA.90.032301,Biswas_2020,Chen2007,sapui2025}. In particular, various quantum correlation measures—including bipartite entanglement~\cite{PhysRevA.66.032110,Osterloh2002}, geometric entanglement~\cite{PhysRevA.71.060305,PhysRevA.93.062341}, generalized geometric measure~\cite{PhysRevA.89.012316,Biswas2014}, and quantum discord~\cite{PhysRevB.78.224413,PhysRevLett.105.095702} have been successfully used to characterize QPTs in a wide range of quantum many-body systems. 
The central role of entanglement in many-body physics is further underscored by its relation to tensor network based numerical methods such as matrix product states~\cite{Klumper_1991,Garcia2007}, density matrix renormalization group~\cite{White1992,Vidal2003,Verstraete2004}, and more general tensor network states~\cite{Niggemann1997,Levin2007,Jiang2008}. Experimental manifestations of these ideas have also been observed in quantum simulators based on optical lattices~\cite{Bloch2008,Nayak2008}, trapped ions~\cite{Leibfried2003,HAFFNER2008155}, photons~\cite{Aspuru-Guzik2012}, and nuclear magnetic resonance systems~\cite{RevModPhys.76.1037}.

More recently, another genuinely quantum resource, namely \emph{nonstabilizerness}, has attracted significant attention in quantum information science. In the stabilizer formalism, quantum states and circuits composed only of stabilizer states and Clifford operations can be efficiently simulated classically, as formalized by the Gottesman--Knill theorem~\cite{gottesman1997stabilizercodesquantumerror,gottesman1998heisenbergrepresentationquantumcomputers,PhysRevA.70.052328,PhysRevLett.116.250501,Bergou2021,Biswas_quanta}. Universal quantum computation, however, requires nonstabilizer resources, collectively referred to as magic~\cite{Veitch_2014,PhysRevLett.118.090501}. Several measures of magic have been proposed, including mana, robustness of magic, and relative entropy of magic, although many of them involve computationally demanding optimizations~\cite{PhysRevA.97.062332,Heinrich2019robustnessofmagic,10.1098/rspa.2019.0251,PhysRevLett.124.090505,PRXQuantum.2.010345,PRXQuantum.3.020333,Dai2022}. Among these, the stabilizer R\'enyi entropy (SRE) has emerged as a computationally accessible and physically meaningful quantifier of nonstabilizer resources~\cite{PhysRevLett.128.050402,Haug2023stabilizerentropies,PhysRevA.110.L040403}.

Although entanglement and magic are distinct resources arising in different operational settings, both are deeply connected to the question of classical simulability. Entanglement constrains the efficiency of tensor-network-based classical simulation methods~\cite{PhysRevLett.91.147902,eisert2013entanglementtensornetworkstates}, while magic quantifies the obstruction to efficient simulation within the stabilizer framework~\cite{Bravyi2019simulationofquantum,deSilva_2024}. This suggests that quantum magic may also encode nontrivial signatures of many-body phenomena, including QPTs. Indeed, recent studies have shown that nonstabilizer-based measures can successfully detect critical behavior in several spin models~\cite{PhysRevA.106.062405,He2023}. Motivated by these developments, it is natural to ask whether nonstabilizer based indicator can serve as an alternative and independent probe of frustrated quantum spin systems, especially in situations where conventional entanglement-based diagnostics exhibit limitations.

The one-dimensional \(J_1\!-\!J_2\) Heisenberg model, where \(J_1\) and \(J_2\) denote the nearest-neighbor and next-nearest-neighbor antiferromagnetic exchange couplings respectively, is a paradigmatic frustrated quantum spin model~\cite{Majumdar1969,White1996,Gu2004,Chhajlany2007}. Its ground-state phase transition from a gapless spin-fluid phase to a gapped dimerized phase has been extensively studied using field-theoretic methods and exact diagonalization, with the critical point located near \(\alpha_c \approx 0.241\), where \(\alpha=J_2/J_1\)~\cite{Haldane1982,Tonegawa1987,OKAMOTO1992433,PhysRevB.54.R9612}. It is known that certain bipartite entanglement measures, such as concurrence~\cite{PhysRevLett.78.5022,PhysRevLett.80.2245}, and other correlations like fidelity, shared purity, nonlocality, and quantum discord can detect this transition when evaluated in the first excited state rather than in the ground state~\cite{Biswas_2020,Chen2007,Biswas_2024,Bao_2024,Bao2026}, while {bipartite entanglement and} multipartite entanglement, such as the generalized geometric measure, may fail to signal the transition in the ground state~\cite{Biswas2014}. In practice, a quantum many-body system can never be prepared exactly at absolute zero temperature, and hence it is unrealistic to assume that the system occupies only its ground state. At low but finite temperatures, the system is more appropriately described by a mixture of the ground state and low-lying excited states. This motivates the investigation of whether information-theoretic order parameters can detect traces of ground-state QPTs in such low-temperature mixed states. In particular, a mixture of the ground state and the first excited state provides a simple but physically meaningful approximation to a thermal state when the temperature is sufficiently low so that higher excited states are negligibly populated. We refer to such a state as the \emph{subjacent state}~\cite{Mondal_2023}. Studying QPT signatures in the subjacent state allows us to probe how robust the critical features are beyond the strict zero-temperature limit. 
{\gb{While the consideration of finite-temperature or mixed states enhances the practical relevance of the analysis, it seemingly conflicts with the established notion that quantum phase transitions are intrinsically zero-temperature phenomena. In this work, we therefore study the nonstabilizer based indicator of both the ground state and the corresponding subjacent mixed state to explore whether this can identify quantum phase transitions that escape detection by conventional quantum correlation measures.}}

In a recent work~\cite{sarkar2026reducedstatestabilizerrenyientropy}, we studied quantum phase transitions in the transverse axial next-nearest-neighbor Ising model and the quantum compass model using the {\gb{purity-corrected}} stabilizer R\'enyi entropy of reduced two-qubit density matrices, and found that {\gb{purity-corrected}} SRE can serve as an efficient local probe of criticality. In the present work, we extend that line of investigation to frustrated Heisenberg-type systems. Specifically, we consider the one-dimensional isotropic \(J_1\!-\!J_2\) Heisenberg model, the one-dimensional anisotropic XXZ \(J_1\!-\!J_2\) model, and the two-dimensional \(J_1\!-\!J_2\) model, and examine whether the second-order purity-corrected SRE of the reduced two-qubit density matrix of the subjacent state can detect the corresponding quantum phase transitions.

The anisotropic \(J_1\!-\!J_2\) model is of particular interest because the anisotropy parameter along the \(z\)-direction enriches the phase structure of the one-dimensional system~\cite{Nomura_1994,Hirata2000,Somma2001}. For \(\alpha \lesssim 0.24\), the system lies in the spin-fluid phase when the anisotropy is less than unity, while for anisotropy greater than unity it enters the N\'eel phase~\cite{Nomura_1994,Hirata2000}. Thus, in addition to the fluid--dimer transition inherited from the isotropic model, the anisotropy introduces further competing orders and phase boundaries. The two-dimensional \(J_1\!-\!J_2\) model is also a canonical frustrated spin system, hosting rich quantum phases due to the competition between nearest-neighbor and next-nearest-neighbor couplings. These models therefore provide a natural testing ground for examining whether nonstabilizer-based local probes can capture frustrated quantum criticality in both one and two spatial dimensions.

{\gb{In this work, we quantify nonstabilizerness through the second-order purity-corrected stabilizer R\'enyi entropy. For an \(n\)-qubit reduced density matrix \(\rho\), it is defined as
\begin{align} \label{Eq_magic}
\widetilde{M_2}(\rho)
&=
-\Bigg[
\log_2\left(
\frac{1}{2^n}
\sum_{P\in\mathcal{P}_n}
\left[\mathrm{Tr}(\rho P)\right]^4
\right)
\nonumber\\
&\qquad
-\log_2\Tr(\rho^2)
\Bigg],
\end{align}
where \(\mathcal{P}_n\) denotes the \(n\)-qubit Pauli group~\cite{PhysRevLett.128.050402,Haug2023stabilizerentropies,PhysRevA.110.L040403}. The second term, \(-\log_2\Tr(\rho^2)\), is the second-order R\'enyi entropy and acts as a purity correction. This correction is necessary because we work with reduced density matrices, which are generally mixed due to entanglement between the subsystem and the traced-out degrees of freedom. By subtracting the mixedness contribution, the quantity \(\widetilde{M_2}\) removes the part arising from entanglement-induced mixedness~\cite{10.21468/SciPostPhys.19.4.085}. Note that the purity-corrected SRE may be interpreted as quantifying nonstabilizerness relative to states with the same degree of mixedness, although it is not a valid magic monotone for mixed states~\cite{10.21468/SciPostPhys.19.4.085}. Rather, for mixed states, the purity-corrected SRE is best understood as a measure of the non-flatness of the Pauli expectation values. Since quantum magic is intimately connected with the non-uniformity of the Pauli expectation-value distribution, the purity-corrected SRE can nevertheless serve as a useful indicator of nonstabilizerness among states of comparable mixedness.

The direct computation of purity-corrected SRE for a full many-body density matrix becomes exponentially expensive with increasing system size, our approach instead {\ab is to} evaluate the purity-corrected SRE of reduced two-qubit density matrices obtained from the ground state and the subjacent state (See subsection~\ref{subsec:subjacent}). This substantially simplifies the computation, as the corresponding Pauli sum contains only \(4^2\) terms, while still retaining essential local information about the many-body state. Such a reduced-density-matrix approach is numerically efficient and enables us to investigate whether local nonstabilizer features can capture signatures of global quantum phase transitions. }}
{Although the purity-corrected SRE can be evaluated for a single-qubit reduced state, we find that it vanishes identically for the one-qubit reduced density matrix of the ground state of the Heisenberg \(J_1\!-\!J_2\) model, both in one-dimension and on the two-dimensional square lattice. By contrast, the corresponding two-qubit reduced density matrices exhibit a nonzero purity-corrected SRE. For the finite systems considered here, the ground state belongs to the total spin singlet sector, as expected for the antiferromagnetic \(J_1\!-\!J_2\) models~\cite{OKAMOTO1992433}. Because a spin singlet is invariant under global SU(2) rotations, the expectation value of every local spin component vanishes,
\(
\langle \sigma_i^x\rangle
=\langle \sigma_i^y\rangle
=\langle \sigma_i^z\rangle
=0.
\)
The single-site reduced density matrix, which can be expressed as
\(
\rho_i
=\frac{1}{2}
\left(
I+
\sum_{\mu=x,y,z}
\langle\sigma_i^\mu\rangle\sigma^\mu
\right),
\)
is therefore maximally mixed, \(\rho_i=I/2\)~\cite{PhysRevA.68.012309}. Since all nonidentity single-qubit Pauli expectation values vanish for this state, its purity-corrected SRE is zero by definition~\cite{PhysRevLett.128.050402}. In contrast, SU(2) symmetry does not require two-body correlators, such as
\(
\left\langle
\boldsymbol{\sigma}_i\!\cdot\!\boldsymbol{\sigma}_j
\right\rangle,
\)
to vanish. These correlators retain information about the short-range singlet, dimer, and frustration-induced correlations of the \(J_1\!-\!J_2\) model, allowing the two-qubit reduced density matrix to exhibit a nonzero purity-corrected SRE.
}

The central goal of this paper is therefore to investigate whether the {\gb{purity-corrected}} SRE of reduced two-qubit density matrices can act as a reliable indicator of QPTs in frustrated \(J_1\!-\!J_2\) spin systems when the many-body system is in a subjacent mixture of its ground and first excited states. By comparing with earlier entanglement-based results, we aim to understand whether nonstabilizer based indicator provides complementary or sharper information about criticality in these models. Since entanglement and nonstabilizerness quantify different aspects of non-classicality, such a comparison may also shed light on the distinct ways in which quantum correlations are redistributed across many-body phase boundaries.

The remainder of the paper is organized as follows. In Sec.~\ref{sec:model}, we introduce the one-dimensional Heisenberg \(J_1\!-\!J_2\) model, the one-dimensional XXZ \(J_1\!-\!J_2\) model, and the two-dimensional \(J_1\!-\!J_2\) model, define the subjacent state as the mixture of the ground and first excited states, describe the reduced two-qubit density-matrix approach used in our numerical analysis and outline our numerical analysis. The behavior of {\gb{purity-corrected}} SRE across the phase transitions of the three models is presented in Sec.~\ref{sec:results}. Finally, we conclude in Sec.~\ref{sec:conclusion} with a summary and discussion of our results.

\section{Models and Method}
\label{sec:model}

In this section, we introduce the frustrated quantum spin models considered in this work, define a low-temperature subjacent state {\ab which has been} used in our analysis {\gb{along with the ground state}}, and briefly describe the reduced-density-matrix approach for evaluating the {\gb{purity-corrected}} stabilizer R\'enyi entropy (SRE).

\subsection{Frustrated \(J_1\!-\!J_2\) spin models}

We study three quantum spin-\(\frac{1}{2}\) models with competing nearest-neighbor and next-nearest-neighbor antiferromagnetic interactions. The competition between these couplings gives rise to frustration and leads to rich quantum phase structures.

\subsubsection{One-dimensional isotropic Heisenberg \(J_1\!-\!J_2\) model}

The Hamiltonian of the one-dimensional isotropic \(J_1\!-\!J_2\) Heisenberg model is given by
\begin{equation}
\label{eq:H_1d_heis}
H_{\mathrm{H}}=
J_1 \sum_{i=1}^{N}\vec{\sigma}_{i}\cdot \vec{\sigma}_{i+1}
+
J_2 \sum_{i=1}^{N}\vec{\sigma}_{i}\cdot \vec{\sigma}_{i+2},
\end{equation}
where \(N\) is the total number of spins, \(J_1>0\) and \(J_2>0\) denote the nearest-neighbor and next-nearest-neighbor antiferromagnetic coupling strengths, respectively, and
\[
\vec{\sigma}_{i}\cdot \vec{\sigma}_{j}
=
\sigma_i^x\sigma_j^x+\sigma_i^y\sigma_j^y+\sigma_i^z\sigma_j^z.
\]
We impose periodic boundary conditions, so that \(\vec{\sigma}_{N+1}=\vec{\sigma}_1\) and \(\vec{\sigma}_{N+2}=\vec{\sigma}_2\). Throughout the paper, we set \(J_1=1\) and use the dimensionless frustration parameter
\begin{equation}
\alpha=\frac{J_2}{J_1}.
\end{equation}

This model exhibits a well-known quantum phase transition from a gapless spin-fluid phase to a gapped dimerized phase as \(\alpha\) is increased, with the critical point located near \(\alpha_c\approx 0.24\) in the thermodynamic limit~\cite{Haldane1982,Tonegawa1987,OKAMOTO1992433,Xu2021,PhysRevB.54.R9612}. At the Majumdar--Ghosh point, \(\alpha=\frac{1}{2}\), the model is exactly solvable and possesses a doubly degenerate dimerized ground state in the thermodynamic limit~\cite{Majumdar1969}.

\subsubsection{One-dimensional XXZ \(J_1\!-\!J_2\) model}

We next consider the anisotropic version of the one-dimensional \(J_1\!-\!J_2\) model, where the interaction in the \(z\)-direction is weighted by an anisotropy parameter \(\delta\). The corresponding Hamiltonian reads
\begin{equation}
\label{eq:H_1d_xxz}
\begin{split}
H_{\mathrm{XXZ}}&=
J_1\sum_{i=1}^{N}
\left(
\sigma_i^x\sigma_{i+1}^x
+\sigma_i^y\sigma_{i+1}^y
+\delta\,\sigma_i^z\sigma_{i+1}^z
\right)\\
&+
J_2\sum_{i=1}^{N}
\left(
\sigma_i^x\sigma_{i+2}^x
+\sigma_i^y\sigma_{i+2}^y
+\delta\,\sigma_i^z\sigma_{i+2}^z
\right),
\end{split}
\end{equation}
again with periodic boundary conditions and \(\alpha=J_2/J_1\). The parameter \(\delta\) controls the strength of the anisotropy along the \(z\)-direction. For \(\delta=1\), Eq.~(\ref{eq:H_1d_xxz}) reduces to the isotropic Heisenberg Hamiltonian in Eq.~(\ref{eq:H_1d_heis}).

The anisotropy enriches the phase diagram of the model. Depending on the values of \(\alpha\) and \(\delta\), the system can host spin-fluid, dimerized, and N\'eel-ordered phases~\cite{Hirata2000,Somma2001,Nomura_1994}. Since the competition between frustration and anisotropy leads to multiple phase boundaries, this model provides an ideal setting for examining whether a local magic-based quantity such as SRE can distinguish different quantum phases.

\subsubsection{Two-dimensional \(J_1\!-\!J_2\) Heisenberg model}

We also consider the frustrated \(J_1\!-\!J_2\) Heisenberg model on a two-dimensional square lattice. Its Hamiltonian is given by
\begin{equation}
\label{eq:H_2d}
H_{\mathrm{2D}}=
J_1\sum_{\langle i,j\rangle}\vec{\sigma}_i\cdot\vec{\sigma}_j
+
J_2\sum_{\langle\!\langle i,j\rangle\!\rangle}\vec{\sigma}_i\cdot\vec{\sigma}_j,
\end{equation}
where \(\langle i,j\rangle\) denotes nearest-neighbor pairs and \(\langle\!\langle i,j\rangle\!\rangle\) denotes next-nearest-neighbor pairs along the diagonals of each plaquette. Here again \(J_1,J_2>0\), and we define the dimensionless frustration parameter as \(\alpha=J_2/J_1\). {\gb{The ground-state phase diagram of this model has been extensively investigated using exact diagonalization and field-theoretical approaches~\cite{richter2010spin,PhysRevLett.80.2705,PhysRevB.79.094413}. Although the precise locations of the phase boundaries remain under active investigation, it is generally accepted that the model hosts two magnetically ordered phases separated by an intermediate quantum paramagnetic phase. As the frustration parameter \(\alpha\) is increased, the system undergoes a transition from the ordinary N\'eel antiferromagnetic phase to the intermediate phase near \(\alpha\approx0.4\), followed by a second transition to the collinear N\'eel phase near \(\alpha\approx0.6\)~\cite{Schulz_1992,PhysRevB.51.6151}. The intermediate phase has been proposed to possess plaquette or columnar dimer order~\cite{PhysRevB.42.8206,PhysRevLett.93.127202,PhysRevB.81.144410,PhysRevB.79.024409,PhysRevB.44.12050,PhysRevB.74.144422,PhysRevB.73.184420,PhysRevLett.91.197202,PhysRevB.78.214415}, while other studies have suggested that it may instead be a quantum spin-fluid phase~\cite{PhysRevB.86.024424,PhysRevB.88.060402,PhysRevB.86.075111}. The ordinary and collinear N\'eel phases are long-range antiferromagnetically ordered phases that exist below the N\'eel temperature~\cite{PhysRevB.86.144411,refId0}.}}

\subsection{Subjacent state: mixture of the ground and first excited states}
\label{subsec:subjacent}

Quantum phase transitions are zero-temperature phenomena. {\gb{We have studied the purity corrected stabilizer R\`enyi entropy in the ground state of the spin models.}}
However, in any realistic setting, a physical many-body system cannot be prepared exactly at absolute zero temperature and is generally susceptible to environmental perturbations. Therefore, rather than assuming that the system occupies only its ground state, it is natural to consider a low-temperature state in which the ground state and the low-lying excited states are populated with nonzero probabilities. {\gb{Earlier studies have shown that, in the one- and two-dimensional frustrated \(J_1-J_2\) spin models, several quantum information measures—including bipartite and multipartite entanglement, shared purity, bipartite and multipartite nonlocality, and fidelity fail to reliably detect quantum phase transitions when evaluated for the ground state alone. Interestingly, these quantities become effective indicators of the phase transitions when evaluated for the first excited state or for a mixed state composed of the ground and first excited states.~\cite{PhysRevA.70.052302,Chen2007,Biswas2014,Mondal_2023,Biswas_2024,Bao_2024}}}

Motivated by the earlier papers, we also consider a mixture of the ground state and the first excited state  of the Hamiltonian. Let \(|{\Psi_0}\rangle\) denote the ground state (ground state is non-degenerate in the considered spin models) and let \(\{|{\Psi_1^{\,i}}\rangle\}_{i=1}^{d}\) denote the orthonormal first excited states, where \(d\) is the degeneracy of the first excited level. We define the \emph{subjacent state}~\cite{Mondal_2023} as
\begin{equation}
\label{eq:subjacent}
\rho
=
(1-p)|{\Psi_0}\rangle\langle{\Psi_0|}
+
\frac{p}{d}\sum_{i=1}^{d}|{\Psi_1^{\,i}}\rangle\langle{\Psi_1^{\,i}}|,
\end{equation}
where \(0\le p\le 1\) is the total occupation probability of the first excited {\ab state}. For \(p=0\), the system is in the ground state, while for small nonzero \(p\), Eq.~(\ref{eq:subjacent}) provides an effective low-temperature approximation in which higher excited states are neglected.

This construction is physically meaningful because, at sufficiently low temperatures, the Boltzmann weights of the ground state and the first excited state(s) dominate over those of higher energy levels. In this sense, the subjacent state captures the leading finite-temperature correction to the ground-state description. Our objective is to investigate whether the signatures of the quantum phase transition {\ab are} visible in the {\gb{purity-corrected}} SRE of reduced two-qubit states obtained from {\gb{the ground state and}} from Eq.~(\ref{eq:subjacent}).

For the numerical calculations, the ground state and the first excited states are obtained by exact diagonalization. In all the cases studied here, we focus on finite-size systems and compute the relevant reduced density matrices directly from the subjacent state in Eq.~(\ref{eq:subjacent}).

In the present work, the mixing probability \(p\) is chosen according to a low-temperature Maxwell--Boltzmann distribution between the ground state and the first excited-state, namely
\begin{equation}
p=
\frac{e^{-\frac{E^{(1)}(N,\alpha)-E^{(0)}(N,\alpha)}{k_B T}}}
{1+e^{-\frac{E^{(1)}(N,\alpha)-E^{(0)}(N,\alpha)}{k_B T}}},
\label{eq:p_boltzmann}
\end{equation}
where \(E^{(0)}(N,\alpha)\) and \(E^{(1)}(N,\alpha)\) denote respectively the ground-state and first-excited-state energies of the system for a given system size \(N\) and frustration parameter \(\alpha\), \(k_B\) is the Boltzmann constant, and \(T\) is the absolute temperature. In our numerical calculations, we fix
\begin{equation}
p = 0.2689 \approx \frac{e^{-1}}{1+e^{-1}},
\label{Eq.(HRI)}
\end{equation}
which corresponds to choosing
\begin{equation}
T \approx \frac{E^{(1)}(N,\alpha)-E^{(0)}(N,\alpha)}{k_B}
\end{equation}
for each value of \(\alpha\). Thus, the subjacent state represents a low-temperature mixture in which the first excited  carries a finite but subdominant weight. The choice in Eq.~(\ref{Eq.(HRI)}) serves as a representative low-temperature setting for illustrating the phase-transition signatures in the models considered here. We note that this particular construction of the subjacent mixed state was introduced earlier in Ref.~\cite{Mondal_2023}.

\subsection{Reduced two-qubit density matrix}

To probe local signatures of quantum criticality, we consider reduced two-qubit density matrices derived from the many-body subjacent state \(\rho\). Let \(a\) and \(b\) denote the two nearest neighbor lattice sites whose reduced state is being studied. The corresponding two-qubit density matrix is obtained by tracing out all the remaining spins:
\begin{equation}
\label{eq:reduced_rho}
\rho_{ab}
=
\Tr_{\overline{ab}}(\rho),
\end{equation}
where \(\Tr_{\overline{ab}}\) denotes the partial trace over every site except \(a\) and \(b\).

Throughout this work, the phrase ``reduced two-qubit density matrix'' refers to the nearest-neighbor two-site reduced state unless explicitly stated otherwise.
This reduced-density-matrix approach is particularly useful because it allows us to investigate whether local properties encode signatures of global phase changes, while keeping the numerical evaluation computationally tractable.

\subsection{Stabilizer R\'enyi entropy of the reduced state}
\label{subsec:sre_method}

We quantify nonstabilizerness through the second-order {\gb{purity-corrected}} stabilizer R\'enyi entropy. For an \(n\)-qubit density matrix \(\varrho\), it is defined as~\cite{PhysRevLett.128.050402,Haug2023stabilizerentropies,PhysRevA.110.L040403}
\begin{equation}
\label{eq:SRE_general}
\widetilde{M_2}(\varrho)
=
-\log_2\left(
\frac{1}{2^n {\gb{\Tr\left(\varrho^2\right)}}}\sum_{P\in\mathcal{P}_n}
\left[\Tr(\varrho P)\right]^4
\right),
\end{equation}
where \(\mathcal{P}_n\) is the \(n\)-qubit Pauli group consisting of all tensor products of single-qubit Pauli operators, including the identity operator. {\gb{The inclusion of purity correction term \(-\log_2\left(\Tr{\varrho^2}\right)\) is described in the introduction section~\ref{intro} after Eq.~(\ref{Eq_magic}). It is defined in reference~\cite{10.21468/SciPostPhys.19.4.085}.}}

In the present work, we evaluate Eq.~(\ref{eq:SRE_general}) for the reduced two-qubit density matrix \(\rho_{ab}\). Since \(n=2\), the Pauli sum runs over only \(4^2=16\) operators, namely
\[
\mathcal{P}_2=\{I,X,Y,Z\}^{\otimes 2}.
\]
Therefore, the two-qubit SRE is given by
\begin{equation}
\label{eq:SRE_2qubit}
\widetilde{M_2}(\rho_{ab})
=
-\log_2\left(
\frac{1}{4{\gb{\Tr\left(\rho_{ab}^2\right)}}}\sum_{P\in\mathcal{P}_2}
\left[\Tr(\rho_{ab}P)\right]^4
\right).
\end{equation}

The local Pauli expectation values entering Eq.~(\ref{eq:SRE_2qubit}) can be extracted from \(\rho_{ab}\). 
By examining the behavior of \(\widetilde{M_2}(\rho_{ab})\) and its derivatives with respect to the control parameters, we analyze whether reduced two-qubit purity-corrected SRE can identify the locations of the underlying phase transitions.

\subsection{Outline of the numerical analysis}

For each model, we vary the frustration parameter \(\alpha=J_2/J_1\), and for the one-dimensional XXZ \(J_1\!-\!J_2\) model we also vary the anisotropy parameter \(\delta\). For every set of Hamiltonian parameters, we obtain the {\gb{ground state and the}} subjacent state in Eq.~(\ref{eq:subjacent}), compute the reduced two-qubit density matrix in Eq.~(\ref{eq:reduced_rho}), and evaluate the corresponding {\gb{purity-corrected}} SRE through Eq.~(\ref{eq:SRE_2qubit}).
The resulting behavior of the reduced two-qubit {\gb{purity-corrected}} SRE is then analyzed as a function of the model parameters in order to identify features associated with quantum criticality.

\section{Quantum phase transition signatures in reduced two-qubit purity-correcetd SRE}
\label{sec:results}

In this section, we present the behavior of the reduced two-qubit stabilizer R\'enyi entropy \(\widetilde{M_2}\) for the three frustrated spin models introduced above. In all cases, the {\gb{purity-corrected}} SRE is evaluated for the {\gb{ground state and the}} subjacent mixed state, i.e., the mixture of the ground state and the first excited-state  with fixed mixing probability \(p=0.2689\). We show that the reduced two-qubit purity-corrected SRE successfully captures the underlying quantum phase transitions in all the models considered here.

\begin{figure}
    \centering
   (a) \includegraphics[width=\linewidth]{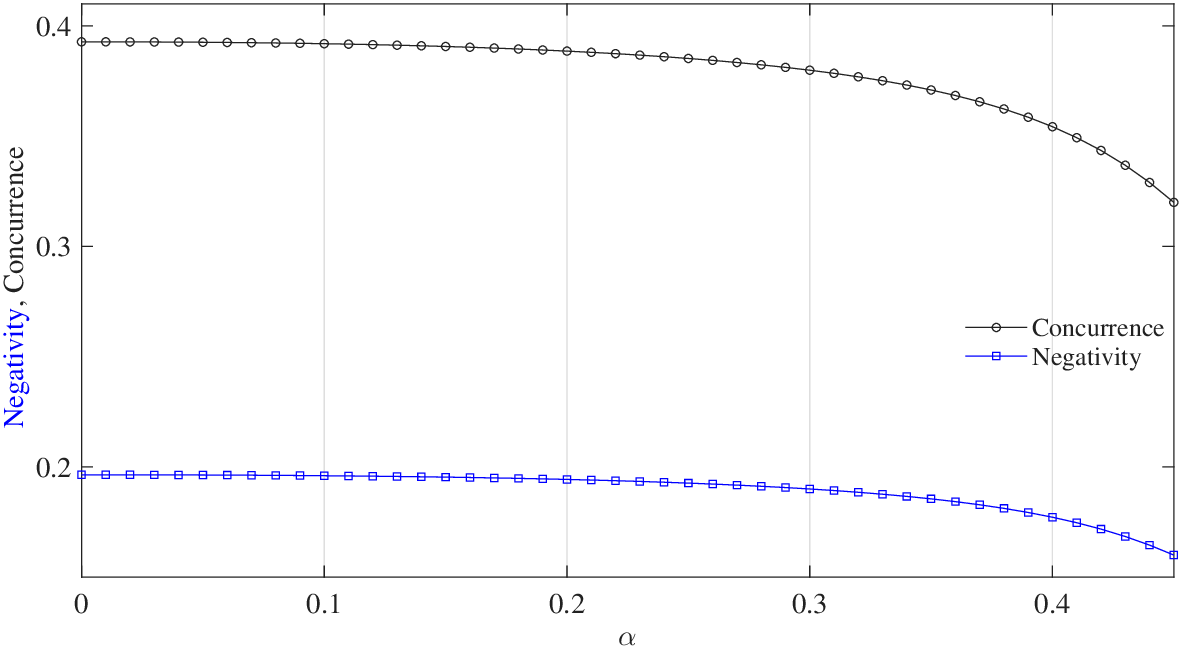}
   
   (b) \includegraphics[width=\linewidth]{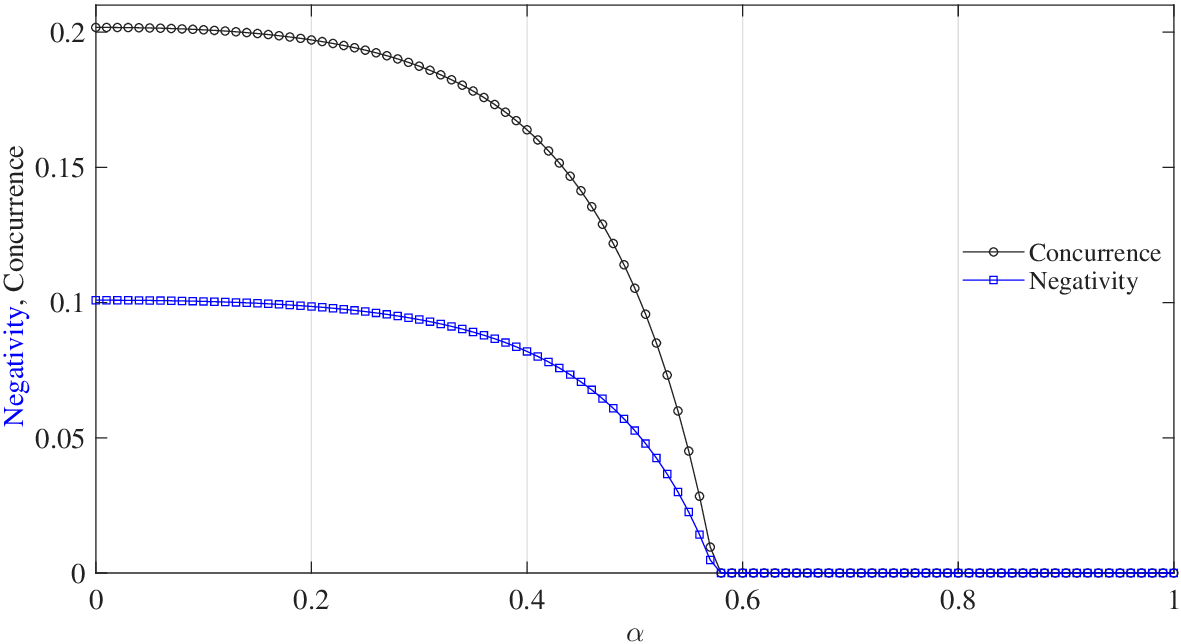}
    \caption{Concurrence and negativity of the reduced two-qubit ground state for the \(J_1\!-\!J_2\) Heisenberg models. (a) One-dimensional model (\(N=16\)). (b) Two-dimensional model (\(N=4\times4\)). In both cases, the bipartite entanglement measures fail to display clear signatures of the quantum phase transitions. In the two-dimensional model, both concurrence and negativity vanish for \(\alpha \gtrsim 0.57\).}
    \label{ground_ent}
\end{figure}

Before presenting the {\gb{purity-corrected}} SRE results, let us briefly recall the known behavior of conventional bipartite entanglement measures for these models. It is already known that, for the one-dimensional and two-dimensional \(J_1\!-\!J_2\) Heisenberg models, bipartite entanglement in the reduced two-qubit ground state does not provide a clear signature of the relevant quantum phase transitions~\cite{Biswas2014,Biswas_2020}. For completeness, in Fig.~\ref{ground_ent}(a) and Fig.~\ref{ground_ent}(b), we plot the concurrence and negativity of the reduced two-qubit ground state for the one-dimensional and two-dimensional \(J_1\!-\!J_2\) Heisenberg models, respectively. In neither case do these quantities exhibit {any signature near the known phase transition points.} In the two-dimensional \(J_1\!-\!J_2\) Heisenberg model, both concurrence and negativity vanish for \(\alpha \gtrsim 0.57\). On the other hand, for the subjacent mixture of the ground state and the first excited state, entanglement is known to exhibit discontinuities near the quantum phase transition points~\cite{Biswas_2020,Mondal_2023}. We will comment on the corresponding behavior in the relevant subsections below.

\subsection{One-dimensional isotropic \(J_1\!-\!J_2\) Heisenberg model}

\begin{figure}
    \centering
    \includegraphics[width=\linewidth]{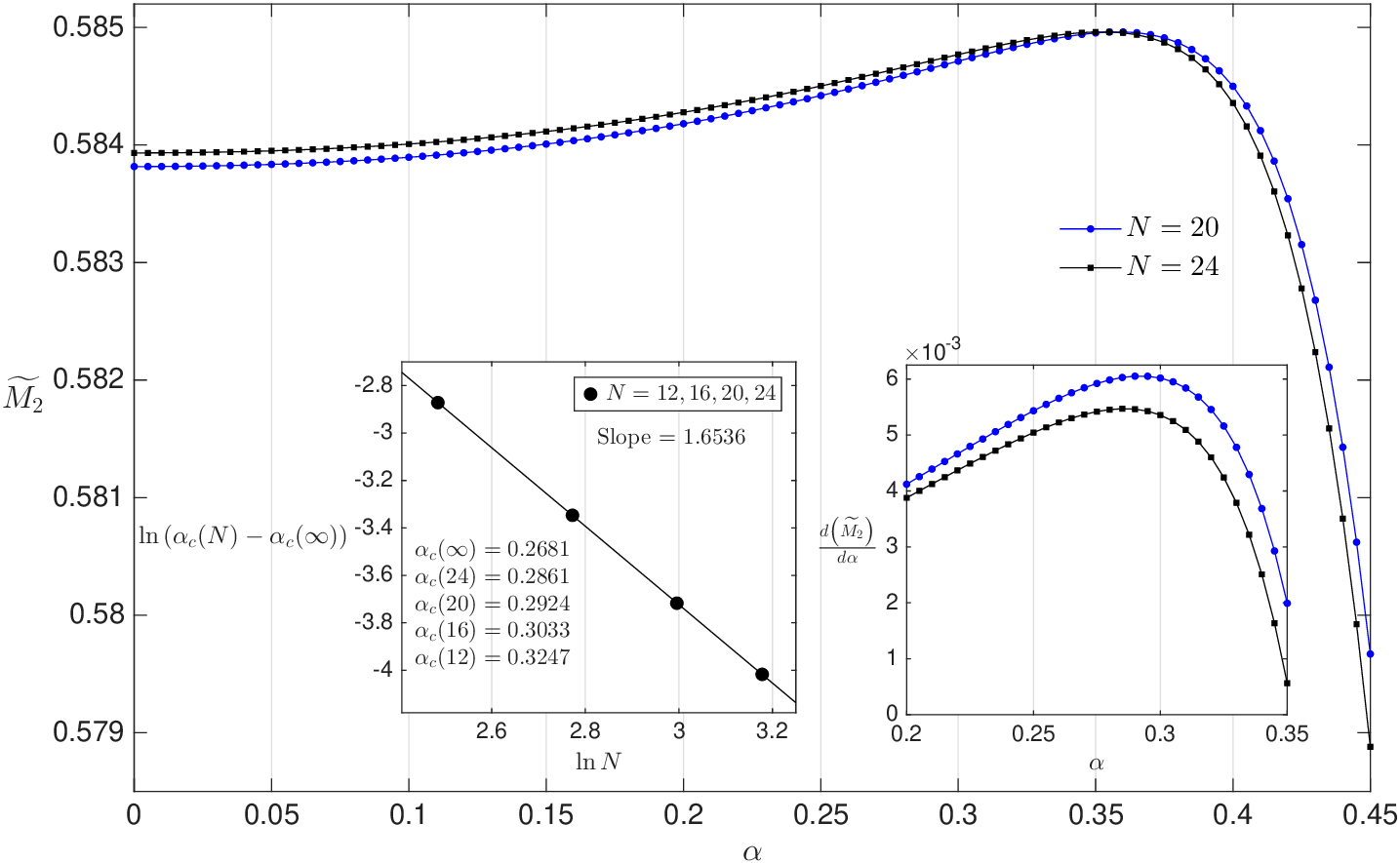}
    \caption{Reduced two-qubit stabilizer R\'enyi entropy \(\widetilde{M_2}\) of the ground state for the one-dimensional isotropic \(J_1\!-\!J_2\) Heisenberg model with system sizes \(N=20,24\). The {\gb{purity-corrected SRE varies smoothly with the frustration parameter \(\alpha\) and exhibits a distinct change in curvature near the known quantum critical point. The right inset shows the corresponding derivative, whose peak identifies the location of the curvature change. The left inset presents the finite-size scaling of the peak positions, \(\alpha_c(N)\), by plotting \(\ln[\alpha_c(N)-\alpha_c(\infty)]\) versus \(\ln N\) for \(N=12,16,20,24\). A linear fit yields the thermodynamic critical point \(\alpha_c(\infty)=0.2681\) and a scaling exponent of \(1.6536\).}}}
    \label{1D_ground}
\end{figure}

We first consider the one-dimensional isotropic \(J_1\!-\!J_2\) Heisenberg chain. As shown in Fig.~\ref{1D_ground}, the reduced two-qubit {\gb{purity-corrected}} SRE of the ground state varies smoothly with the frustration parameter \(\alpha\), {\gb{while exhibiting a distinct change in curvature in the vicinity of the quantum critical point. This suggests that the local nonstabilizer resource grows with an increasing rate throughout the spin-fluid phase. Close to the onset of the dimerized phase, however, this growth trend changes qualitatively: although the reduced purity-corrected SRE continues to increase over a narrow range, its growth rate decreases, producing {the point of inflection in the plot.} The reduced purity-corrected SRE subsequently reaches a maximum before decreasing further inside the dimerized phase. This feature provides a clear signature of the quantum phase transition. Thus, unlike conventional bipartite entanglement measures, which fail to detect the transition from the ground state alone, the reduced two-qubit purity-corrected SRE successfully identifies the critical point in this model.}}
{\gb{To estimate the critical point in the thermodynamic limit, we perform a finite-size scaling analysis using system sizes \(N=12,16,20,24\). The left inset of Fig.~\ref{1D_ground} shows a linear fit of
\[
\ln\!\left[\alpha_c(N)-\alpha_c(\infty)\right]
\ {versus} \
\ln N,
\]
from which we obtain
\[
\alpha_c(\infty)= {\ab 0.2681}.
\]
The individual finite-size estimates used in the fit are shown in the left inset of Fig.~\ref{1D_ground}.
The fitted slope is \(1.6536\), indicating the scaling relation
\[
\alpha_c(N)-\alpha_c(\infty) \approx N^{-1.6536} + {constant}.
\] 
The widely reported critical value is near \(\alpha \approx 0.241\), and our estimate is \(\alpha \approx 0.2681\).}}

\begin{figure}
    \centering
    \includegraphics[width=\linewidth]{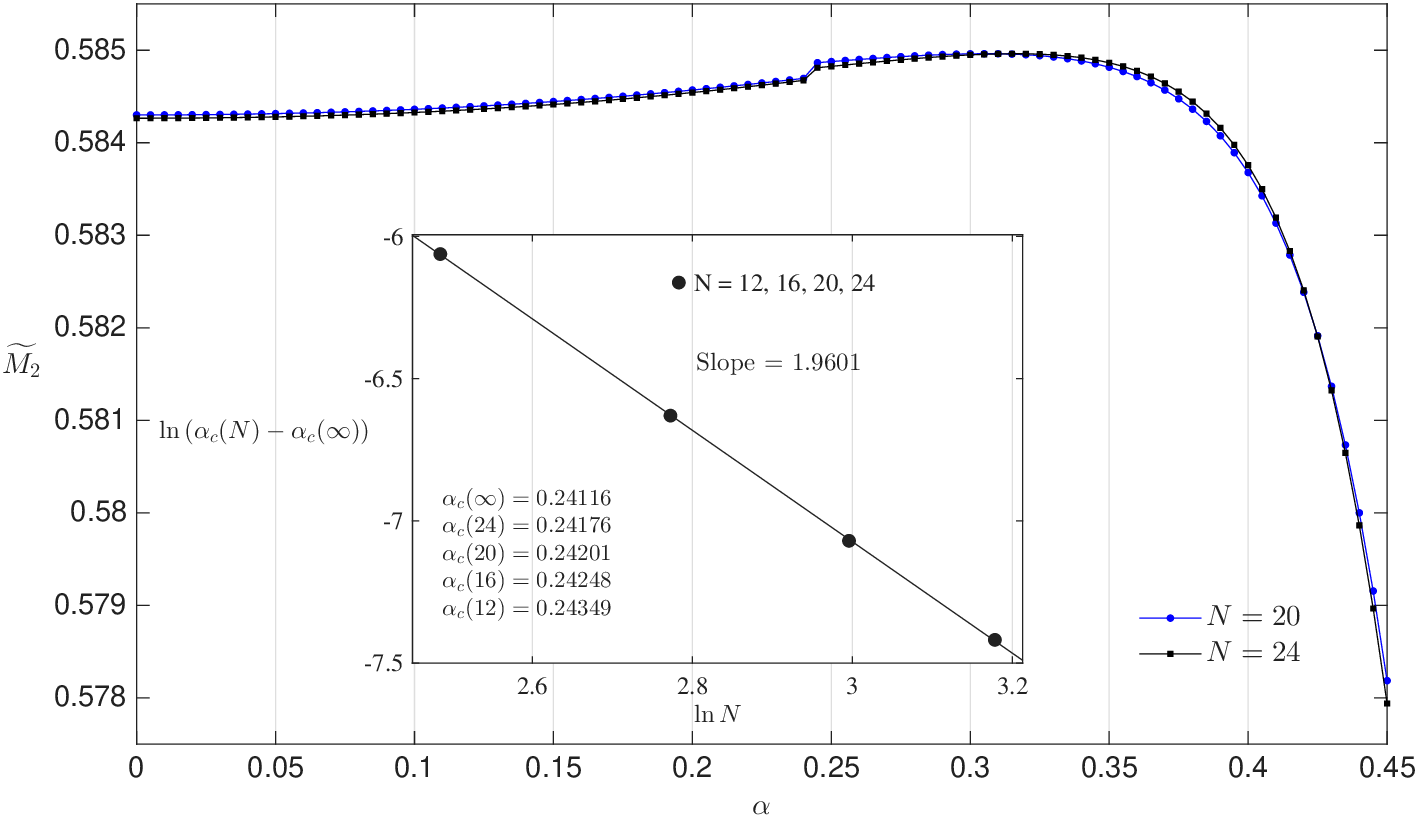}
    \caption{Reduced two-qubit stabilizer R\'enyi entropy \(\widetilde{M_2}\) of the subjacent mixture of the ground state and the first excited-state for the one-dimensional isotropic \(J_1\!-\!J_2\) Heisenberg model as a function of the frustration parameter \(\alpha\), for \(N=20\) and \(N=24\). In both cases, \(\widetilde{M_2}\) exhibits a discontinuity near the quantum critical point. The inset shows the finite-size scaling of \(\ln[\alpha_c(N)-\alpha_c(\infty)]\) versus \(\ln N\) for \(N=12,16,20,24\), yielding \(\alpha_c(\infty)=0.24116\) and slope \(1.9601\).}
    \label{1D}
\end{figure}

We {\gb{then}} turn to the subjacent state, i.e., the mixture of the ground state and the first excited-state  with mixing probability \(p=0.2689\).
The corresponding behavior of \(\widetilde{M_2}\) as a function of the frustration parameter \(\alpha=J_2/J_1\) is shown in Fig.~\ref{1D} for system sizes \(N=20\) and \(N=24\). In both cases, the {\gb{purity-corrected}} SRE increases monotonically with \(\alpha\), but more importantly it exhibits a clear discontinuity in a narrow region near the known critical point. This {discontinuity} signals the transition from the gapless spin-fluid phase to the dimerized phase.
For \(N=20\), the discontinuity is located at
\(
\alpha_c(20)=0.24201,
\)
while for \(N=24\), it occurs at
\(
\alpha_c(24)=0.24176.
\)
Thus, the critical point drifts toward smaller \(\alpha\) as the system size increases.

To estimate the critical point in the thermodynamic limit, we perform a finite-size scaling analysis using system sizes \(N=12,16,20,24\). The inset of Fig.~\ref{1D} shows a linear fit of
\[
\ln\!\left[\alpha_c(N)-\alpha_c(\infty)\right]
\ {versus} \
\ln N,
\]
from which we obtain
\[
\alpha_c(\infty)=0.24116.
\]
The individual finite-size estimates used in the fit are shown in the inset of Fig.~\ref{1D}.
The fitted slope is \(1.9601\), indicating the scaling relation
\[
\alpha_c(N)-\alpha_c(\infty) \approx N^{-1.9601} + {constant}.
\]
The value \(\alpha_c(\infty)=0.24116\) is in excellent agreement with the known critical point of the model. Therefore, the reduced two-qubit {\gb{purity-corrected}} SRE of the subjacent state successfully detects the spin-fluid to dimer transition and reproduces the {known} thermodynamic-limit critical point. 

\begin{figure}
    \centering
    \includegraphics[width=0.95\linewidth]{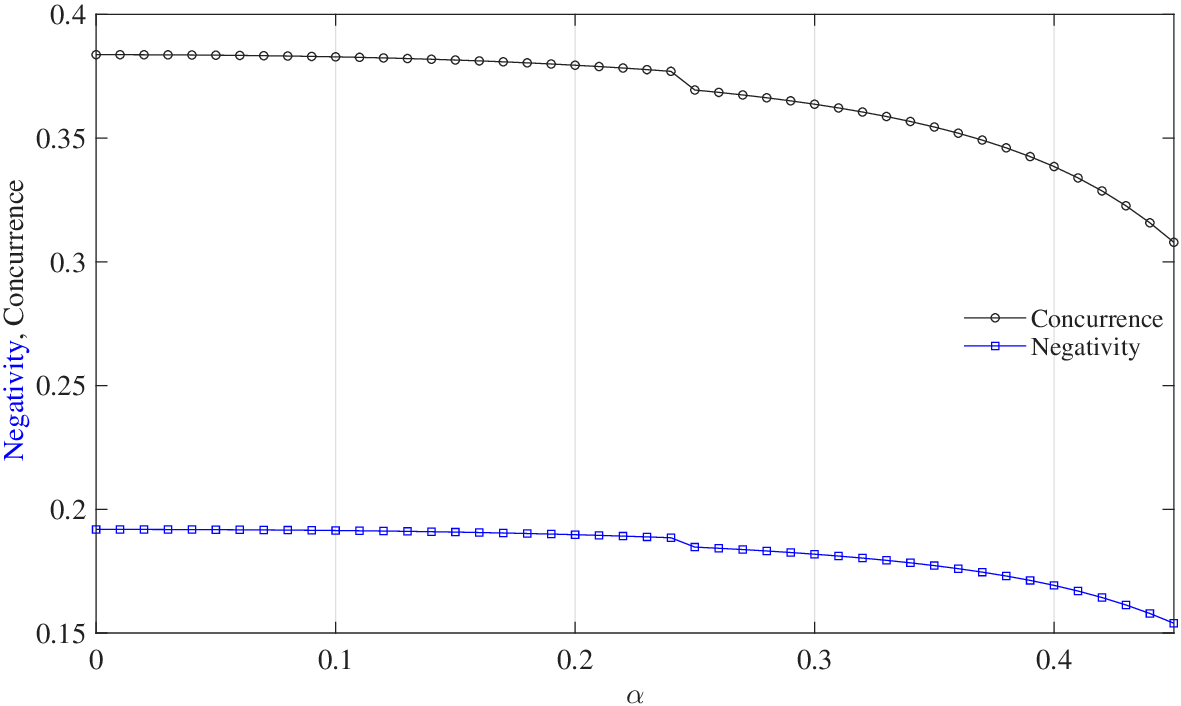}
    \caption{Reduced two-qubit concurrence and negativity of the subjacent mixture of the ground state and the first excited-state  for the one-dimensional isotropic \(J_1\!-\!J_2\) Heisenberg model with \(N=16\). Both quantities exhibit clear nonanalytic signatures near the known quantum critical point.}
    \label{1D_con_ngtvy}
\end{figure}

It is worth noting that the bipartite entanglement of the first excited state, as well as that of the subjacent mixed state composed of the ground and first excited states, is known to exhibit clear signatures of the quantum phase transition in this model~\cite{Biswas_2020,Mondal_2023}. For completeness, Fig.~\ref{1D_con_ngtvy} shows the reduced two-qubit concurrence and negativity for the subjacent mixed state with mixing probability \(p=0.2689\). The entanglement measures clearly identify the critical point, in agreement with previous studies.
{\gb{Thus, for the subjacent mixed state, reduced two-qubit purity-corrected SRE is found to be as effective as reduced two-qubit bipartite entanglement in detecting the quantum phase transition. 
The distinction, however, arises when only the ground state is considered. In that case, reduced two-qubit purity-corrected SRE exhibits a clear signature of the quantum phase transition through a change in curvature, whereas the bipartite entanglement measures fail to detect the transition (see Fig.~\ref{ground_ent}~(a)). In this sense, reduced two-qubit purity-corrected SRE not only complements bipartite entanglement but also provides a more sensitive indicator of quantum criticality in the one-dimensional \(J_1\!-\!J_2\) Heisenberg model. }} 

{\gb{One possible interpretation of the observed curvature change of the purity corrected SRE in the ground state is the following. Within the spin-fluid phase, increasing frustration enhances quantum fluctuations and the competition among different valence-bond configurations. As a consequence, the local reduced state moves progressively farther from the stabilizer manifold, leading to an accelerating growth of the reduced two-qubit purity-corrected SRE. Beyond the critical point, however, the system enters the gapped dimerized phase, where the dominant local structure is a collection of nearest-neighbor singlets. Further increasing the frustration primarily reinforces this dimer order rather than generating qualitatively new local nonstabilizer features. Consequently, the growth of the reduced purity-corrected SRE slows down, giving rise to the observed change from concave to convex behavior. Although this interpretation is physically appealing, establishing a direct connection between dimerization and the redistribution of nonstabilizer resources remains an interesting open problem.}}

\subsection{One-dimensional XXZ \(J_1\!-\!J_2\) model}

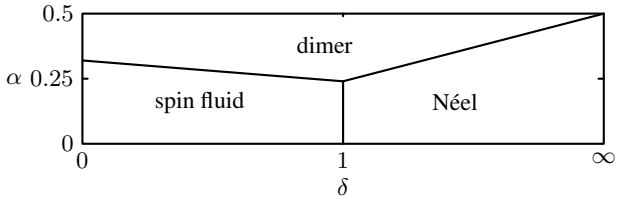
\begin{figure}[ht]
\centering
\begin{tikzpicture}[scale=3.45, line cap=round, line join=round]

\draw[thick] (0,0) rectangle (2,0.5);

\draw[thick] (0,0.32) -- (1,0.24);
\draw[thick] (1,0.24) -- (2,0.5);
\draw[thick] (1,0) -- (1,0.24);

\draw[thick] (0,0.25) -- (0.02,0.25);
\draw[thick] (2,0.25) -- (1.98,0.25);
\draw[thick] (1,0.5) -- (1,0.49);

\node[left] at (0,0) {$0$};
\node[left] at (0,0.25) {$0.25$};
\node[left] at (0,0.5) {$0.5$};

\node[below] at (0,0) {$0$};
\node[below] at (1,0) {$1$};
\node[below] at (2,0) {$\infty$};

\node[left] at (-0.2,0.25) {$\alpha$};
\node[below] at (1,-0.1) {$\delta$};

\node at (0.45,0.16) {spin fluid};
\node at (0.93,0.38) {dimer};
\node at (1.43,0.16) {N\'eel};

\end{tikzpicture}
\caption{{\gb{Schematic phase diagram of the one-dimensional XXZ \(J_1\!-\!J_2\) model. Known from existing literature~\cite{10.1143/PTPS.145.113}}}}
\label{fig:xxz_phase_diagram}
\end{figure}

We now turn to the anisotropic XXZ \(J_1\!-\!J_2\) chain, where the anisotropy parameter \(\delta\) controls the interaction strength in the \(z\)-direction. 
{\gb{Fig.~\ref{fig:xxz_phase_diagram} shows the known schematic phase diagram of the one-dimensional XXZ \(J_1\!-\!J_2\) spin model. Depending on the frustration parameter \(\alpha=J_2/J_1\) and the anisotropy parameter \(\delta\), the model exhibits three distinct phases: the spin-fluid, dimerized, and N\'eel phases. For \(\delta=1\), the Hamiltonian reduces to the isotropic one-dimensional \(J_1\!-\!J_2\) Heisenberg model discussed in the previous subsection. We first investigate the reduced two-qubit purity-corrected SRE of the ground state for different values of the anisotropy parameter \(\delta\).}}

\begin{figure}
    \centering
    \includegraphics[width=\linewidth]{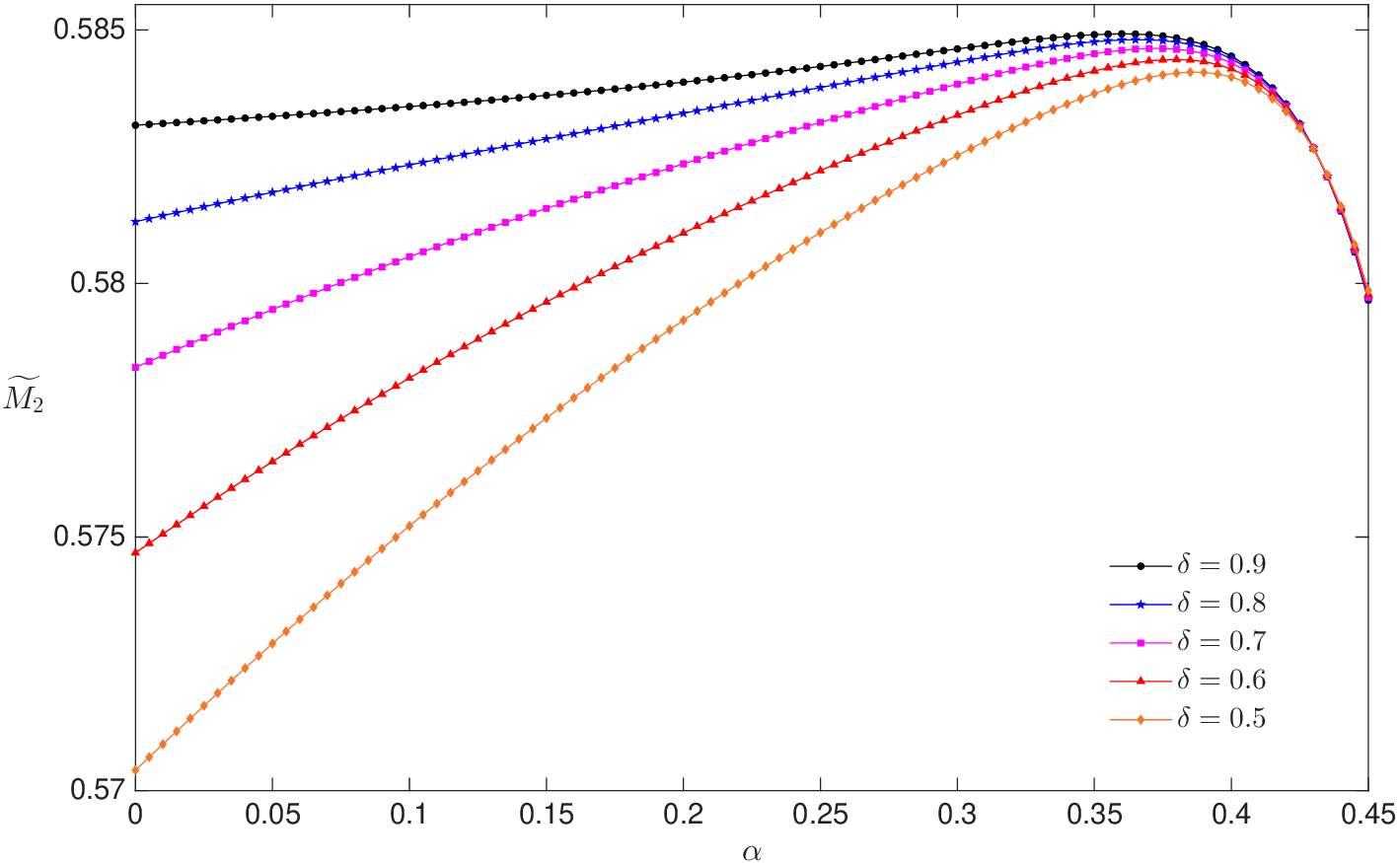}
    
    (a)
    
    \includegraphics[width=\linewidth]{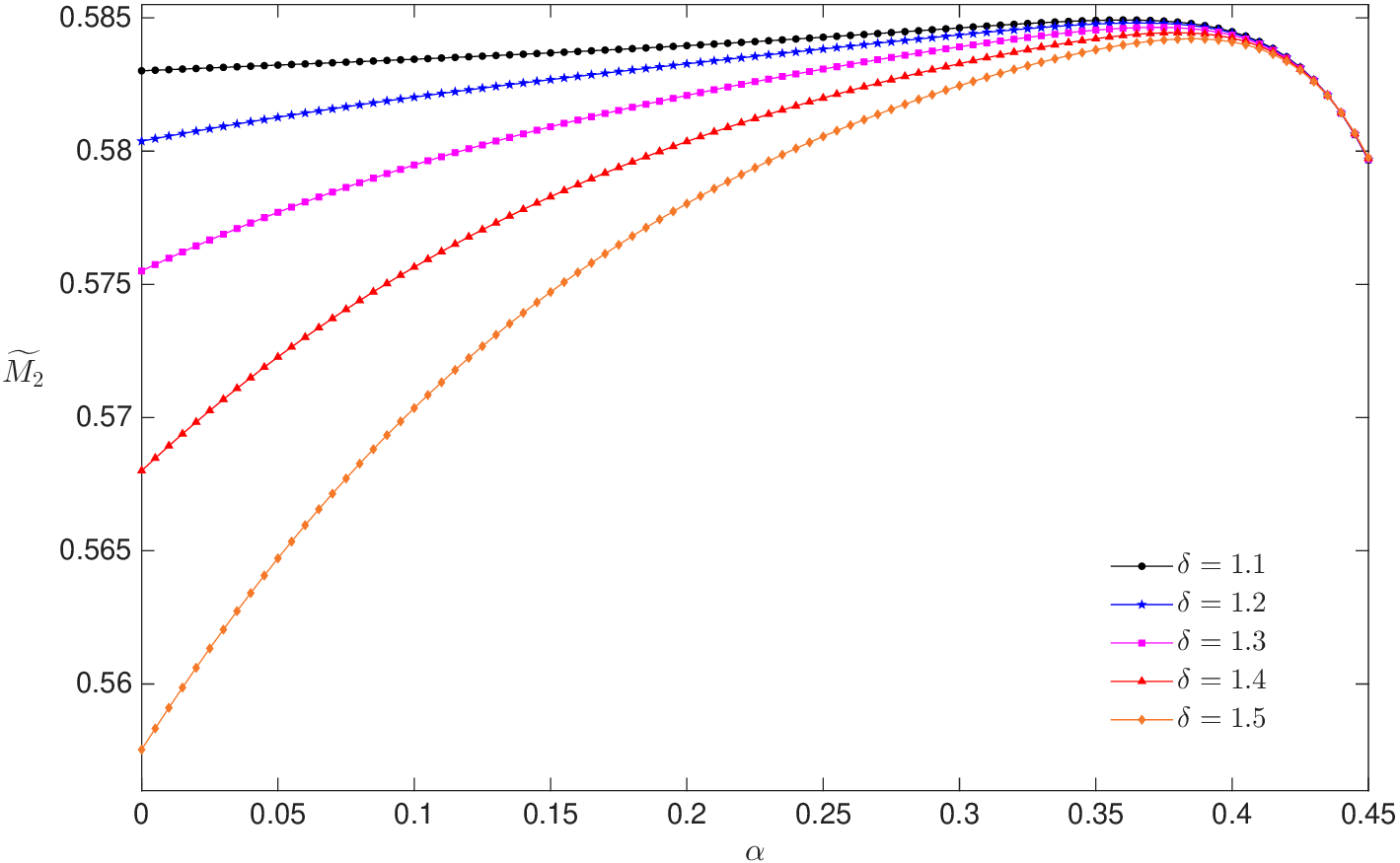}
    
    (b)
    \caption{{\gb{Behavior of the reduced two-qubit stabilizer R\'enyi entropy \(\widetilde{M_2}\) of the ground state for the one-dimensional XXZ \(J_1\!-\!J_2\) model with \(N=20\). (a) \(\delta<1\). (b) \(\delta>1\).}}}
    \label{1D_delta_ground}
\end{figure}

{\gb{As shown in Fig.~\ref{1D_delta_ground}, when the anisotropy deviates only slightly from the isotropic point (\(\delta=1\)), the reduced two-qubit SRE continues to exhibit a change in curvature from concave to convex near the spin-fluid--dimer transition {\ab and the N\'eel-dimer transition}. However, as the anisotropy is increased further away from the isotropic point, this {inflection point} signature disappears, and the SRE remains convex throughout the parameter range. Consequently, the reduced two-qubit purity-corrected SRE of the ground state cannot be regarded as a reliable indicator of the spin-fluid--dimer and dimer-N\`eel quantum phase transitions throughout the XXZ \(J_1\!-\!J_2\) phase diagram.}}

{\gb{For \(\alpha \leq 0.24\), the system undergoes a transition from the spin-fluid phase to the N\'eel phase as the anisotropy parameter \(\delta\) is varied across the isotropic point \(\delta=1\). To investigate whether the reduced two-qubit purity-corrected SRE captures this transition in the ground state, we study the behavior of the purity-corrected SRE as a function of \(\delta\) for representative values \(\alpha=0.15\) and \(0.2\). The results are shown in Fig.~\ref{delta_vary_ground}. In both cases, the reduced two-qubit SRE attains its maximum at the isotropic point \(\delta=1\) and decreases as \(\delta\) is varied away from this point in either direction. 

\begin{figure}
    \centering
    \includegraphics[width=\linewidth]{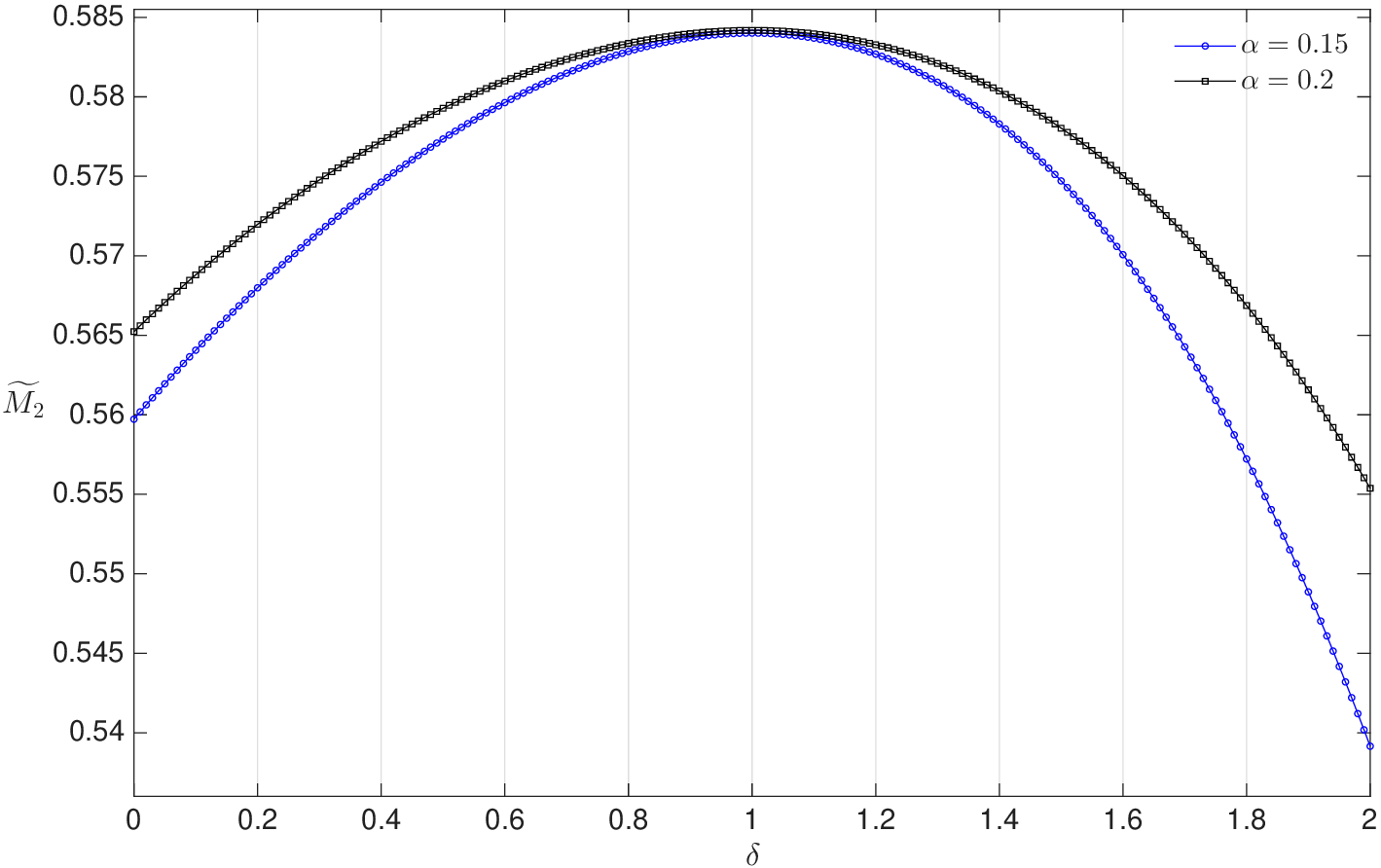}
    \caption{{\gb{Behavior of the reduced two-qubit purity-corrected stabilizer R\'enyi entropy \(\widetilde{M_2}\) of the ground state for the one-dimensional XXZ \(J_1\!-\!J_2\) model with \(N=20\) as a function of the anisotropy parameter \(\delta\), for \(\alpha=0.15\) and \(0.2\). In both cases, \(\widetilde{M_2}\) attains its maximum at the isotropic point \(\delta=1\) and decreases smoothly as \(\delta\) is varied away from unity.}}}
    \label{delta_vary_ground}
\end{figure}

{\gb{We then studied reduced purity-corrected SRE in the subjacent mixed state.}}
Figure~\ref{1D_delta}(a) shows the variation of \(\widetilde{M_2}\) {\ab{ of the subjacent state}} as a function of both \(\alpha\) and \(\delta\) for \(N=20\). The resulting surface clearly exhibits {discontinuities} that trace out the phase boundaries in the \((\alpha,\delta)\) plane.

\begin{figure}
    \centering
    \includegraphics[width=\linewidth]{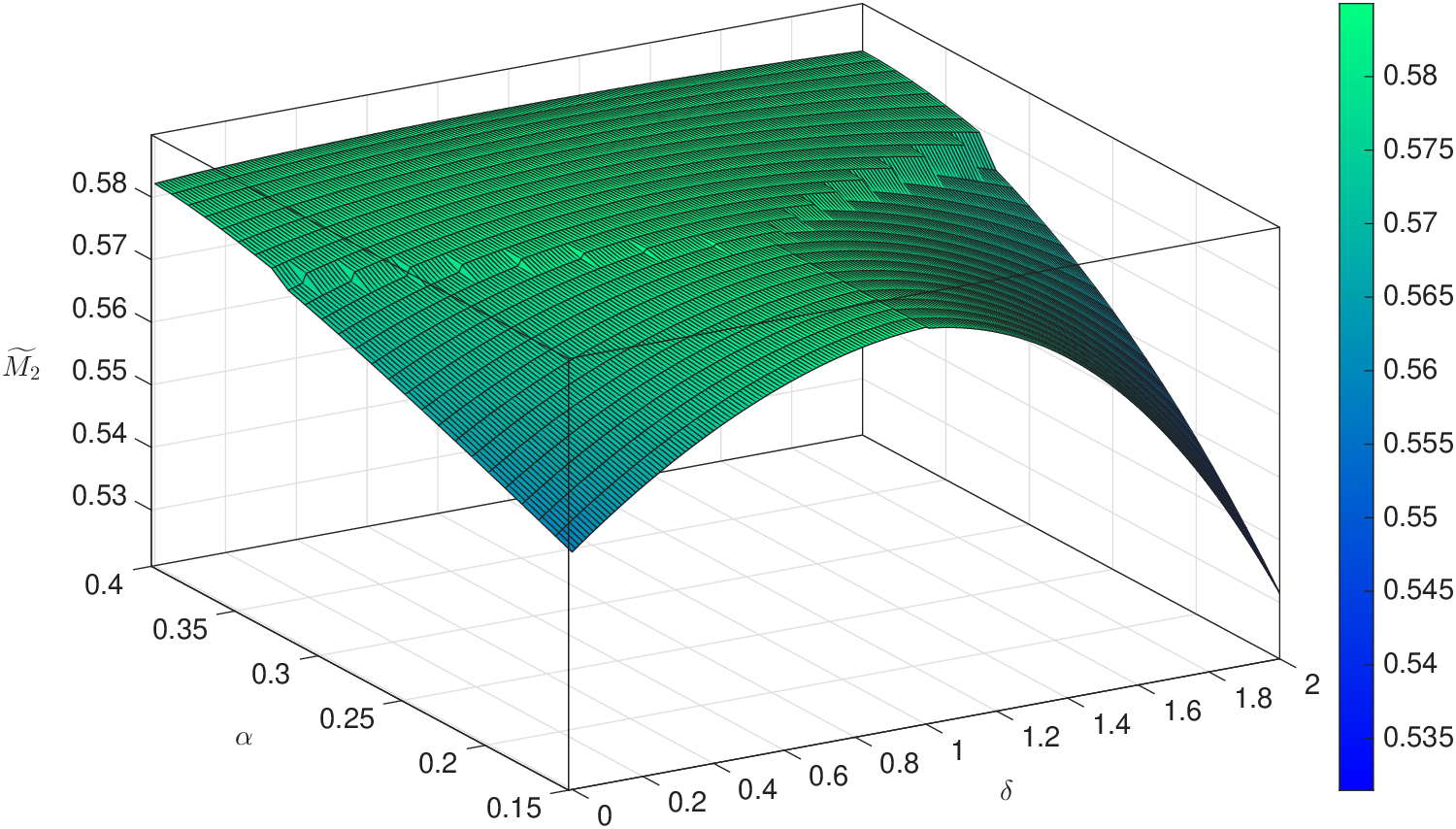}
(a)
    \includegraphics[width=\linewidth]{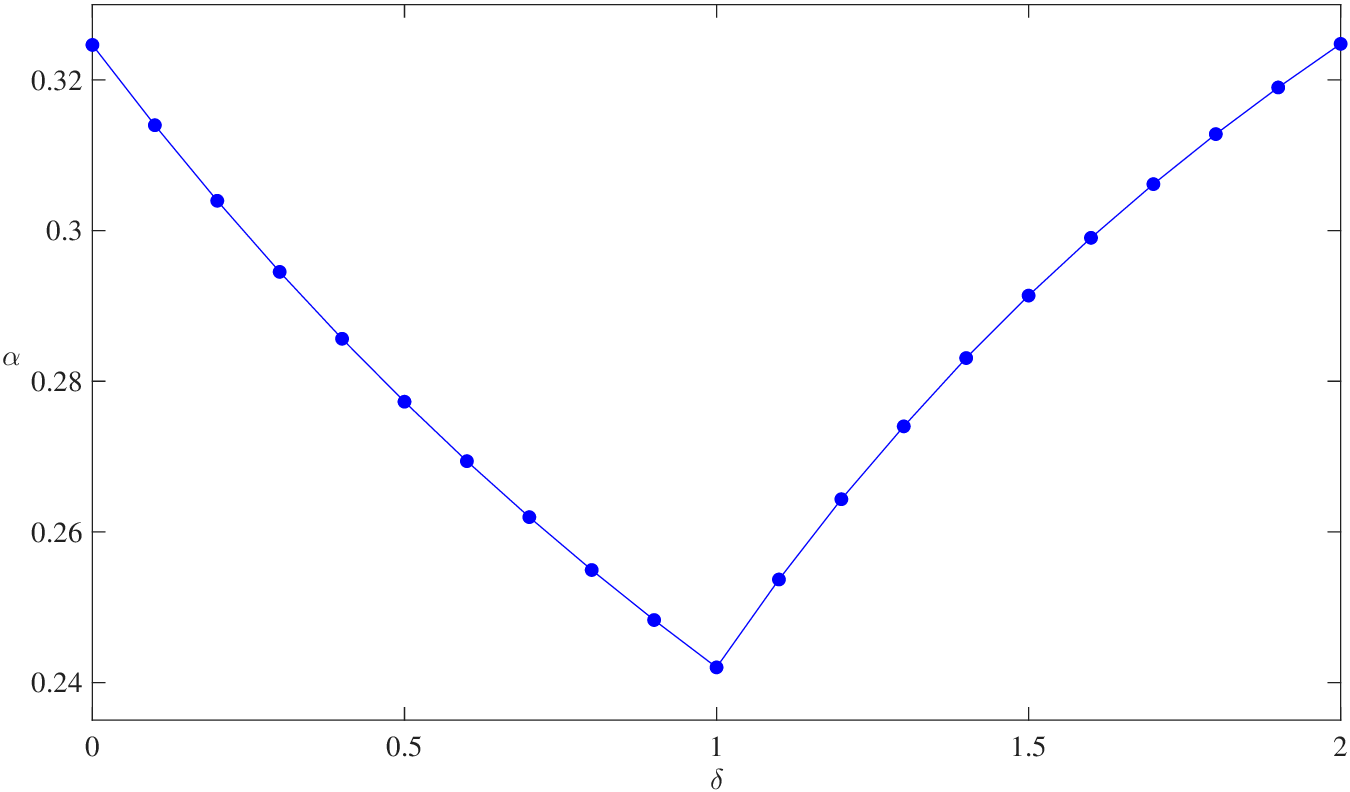}
(b)
    \caption{Behavior of the reduced two-qubit stabilizer R\'enyi entropy \(\widetilde{M_2}\) of the subjacent mixture of the ground state and the first excited-state for the one-dimensional XXZ \(J_1\!-\!J_2\) model with \(N=20\). (a) Surface plot of \(\widetilde{M_2}\) as a function of the frustration parameter \(\alpha\) and the anisotropy parameter \(\delta\). The step-like nonanalytic structure visible on the surface marks the phase boundary for \(\alpha>\alpha_c(\delta)\). (b) Phase boundary extracted from the nonanalytic points of \(\widetilde{M_2}\) for \(\delta=0,0.1,\dots,2\). The critical line has a V-shaped profile with minimum near \((\delta,\alpha)=(1,0.242)\).}
    \label{1D_delta}
\end{figure}

\begin{figure}
    \centering
    \includegraphics[width=\linewidth]{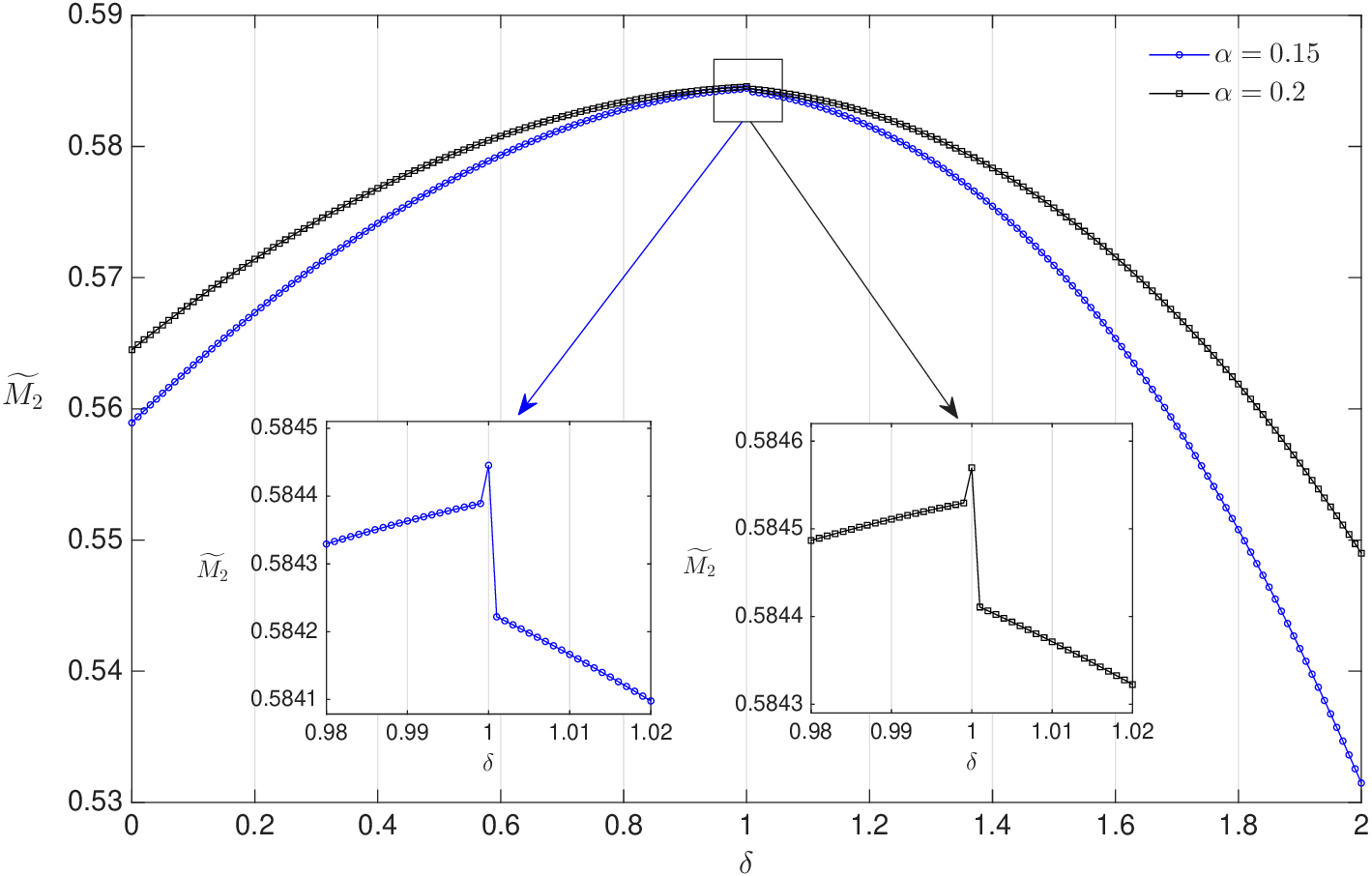}
    \caption{Behavior of the reduced two-qubit stabilizer R\'enyi entropy \(\widetilde{M_2}\) of the subjacent mixture of the ground state and the first excited-state for the one-dimensional XXZ \(J_1\!-\!J_2\) model with \(N=20\) as a function of \(\delta\) for \(\alpha=0.15\) and \(\alpha=0.2\), both lying below the isotropic critical value. In both cases, the transition at \(\delta=1\) is detected through a sharp kink, as shown in the magnified insets.}
    \label{1D_delta_alpha_less_than}
\end{figure}

For \(\alpha>0.24\), the transition is detected through a discontinuity in \(\widetilde{M_2}\), visible on the surface as a sharp step-like feature. The corresponding phase boundary extracted from these discontinuities is plotted in Fig.~\ref{1D_delta}(b). The phase boundary has a pronounced V-shaped form, attaining its minimum near the isotropic point \(\delta=1\), where
\(
\alpha_c \approx 0.242.
\)
As \(\delta\) moves away from unity, the critical value of \(\alpha\) increases on both sides. In particular, from Fig.~\ref{1D_delta}(b), \(\alpha_c\) rises from approximately \(0.242\) at \(\delta=1\) to about \(0.325\) at \(\delta=0\) and to about \(0.325\) at \(\delta=2\). This behavior is consistent with the fact that anisotropy shifts the fluid--dimer boundary~\cite{10.1143/PTPS.145.113}.

Fig.~\ref{1D_delta_alpha_less_than} shows \(\widetilde{M_2}\) as a function of \(\delta\) for two representative values, \(\alpha=0.15\) and \(\alpha=0.2\). In both cases, \(\widetilde{M_2}\) {has a discontinuity} at
\(
\delta=1,
\)
marking the transition across the isotropic line. The magnified insets show that this feature appears not as a broad extremum but as a kink-like singularity localized at \(\delta=1\). For \(\delta<1\), \(\widetilde{M_2}\) increases gradually toward the isotropic point, while for \(\delta>1\), it decreases. Thus, even when the frustration strength is below the isotropic critical value, the reduced two-qubit {\gb{purity-corrected}} SRE remains sensitive to the anisotropy-driven phase boundary.

Taken together, Figs.~\ref{1D_delta}(a),~\ref{1D_delta}(b),~and~\ref{1D_delta_alpha_less_than} show that the reduced two-qubit {\gb{purity-corrected}} SRE {\gb{of the subjacent mixed state}} provides a unified description of the phase diagram of the XXZ \(J_1\!-\!J_2\) chain. The extracted phase boundary reproduces the expected V-shaped structure with minimum near \((\delta,\alpha)=(1,0.242)\).

\subsection{Two-dimensional \(J_1\!-\!J_2\) Heisenberg model}

\begin{figure}
    \centering
    \includegraphics[width=\linewidth]{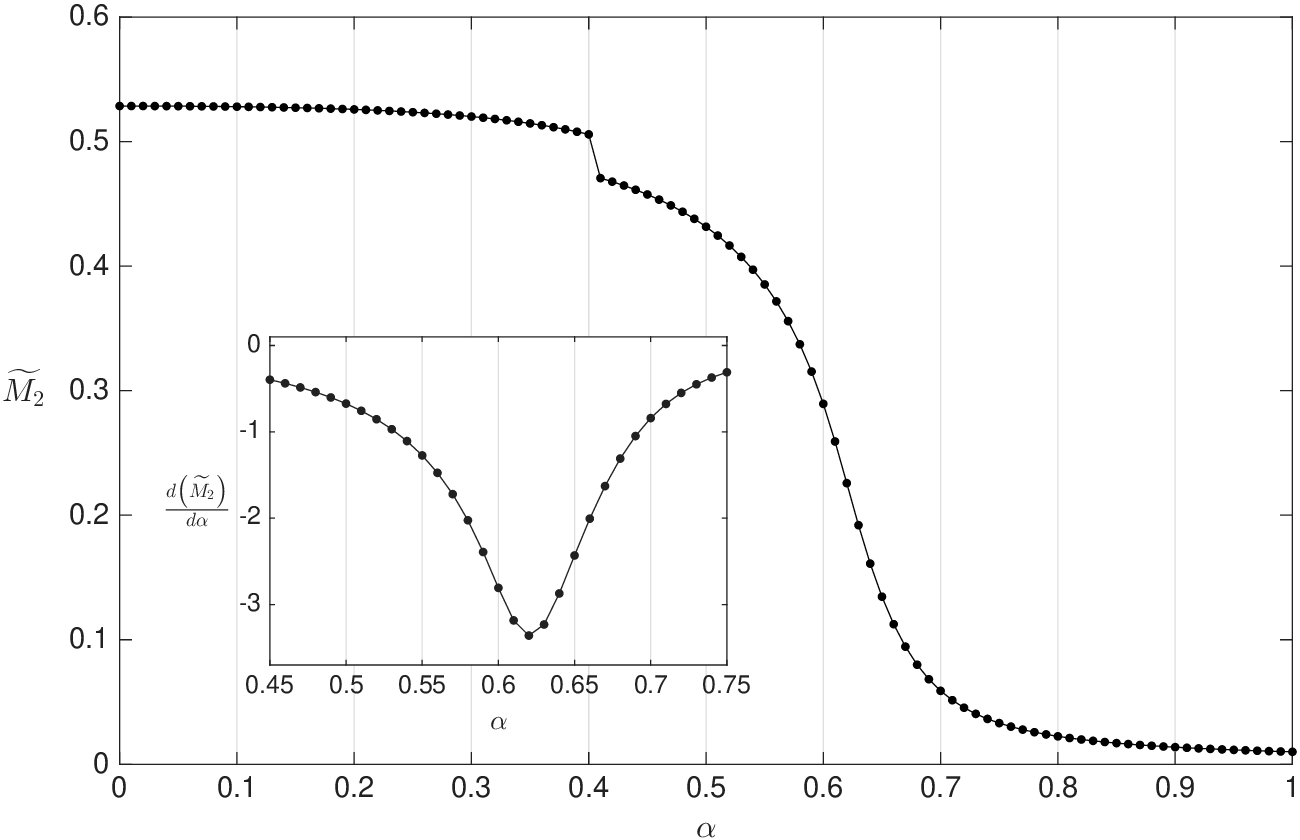}
    \caption{Reduced two-qubit {\gb{purity-corrected}} stabilizer R\'enyi entropy \(\widetilde{M_2}\) of the subjacent mixture of ground state and first excited state for the two-dimensional \(J_1\!-\!J_2\) Heisenberg model on a \(4\times4\) square lattice (\(N=16\)) with periodic boundary conditions. A discontinuity at \(\alpha\approx 0.40781\) signals the first phase transition. A second transition near \(\alpha\approx 0.6208\) is identified through a change in curvature of \(\widetilde{M_2}\); the inset shows the derivative \(d(\widetilde{M_2})/d\alpha\), whose minimum locates the transition.}
    \label{2D}
\end{figure}

We finally consider the two-dimensional \(J_1\!-\!J_2\) Heisenberg model on a \(4\times 4\) square lattice with periodic boundary conditions. The reduced two-qubit {\gb{purity-corrected}} SRE of the subjacent mixture of ground state and first excited state, is shown in Fig.~\ref{2D} as a function of \(\alpha\).

The behavior of \(\widetilde{M_2}\) reveals two distinct quantum phase transitions. The first transition occurs near
\(
\alpha \approx 0.40781,
\)
where the {\gb{purity-corrected}} SRE displays a clear discontinuity. {This} discontinuity signals the transition from the ordinary N\'eel phase to an intermediate frustrated phase.

As \(\alpha\) is increased further, \(\widetilde{M_2}\) continues to decrease smoothly, but its curvature changes near
\(
\alpha \approx 0.6208.
\)
This second transition is not accompanied by a discontinuity in \(\widetilde{M_2}\) itself. Instead, it is identified through the behavior of the derivative \(d(\widetilde{M_2})/d\alpha\), shown in the inset of Fig.~\ref{2D}. The derivative decreases with \(\alpha\), reaches a minimum near \(\alpha\approx 0.6208\), and then increases for larger \(\alpha\). Thus, in the two-dimensional model, the first transition is detected through a discontinuity in \(\widetilde{M_2}\), while the second is captured through a {point of inflection}. Overall, the reduced two-qubit {\gb{purity-corrected}} SRE successfully detects both phase boundaries in the two-dimensional frustrated model: the transition near \(\alpha\approx 0.40781\), associated with the onset of an intermediate phase, and the transition near \(\alpha\approx 0.6208\), associated with the crossover from the intermediate phase to the collinear N\'{e}el phase.

It is known that bipartite entanglement of the first excited state alone, as well as that of the subjacent mixture of the ground state and the first excited state, can detect the first quantum phase transition near \(\alpha \approx 0.4\) in this model~\cite{Biswas_2020,Mondal_2023}. For completeness, in Fig.~\ref{2D_con_ngtvy}, we plot the reduced two-qubit concurrence and negativity of the subjacent state.

\begin{figure}
    \centering
    \includegraphics[width=\linewidth]{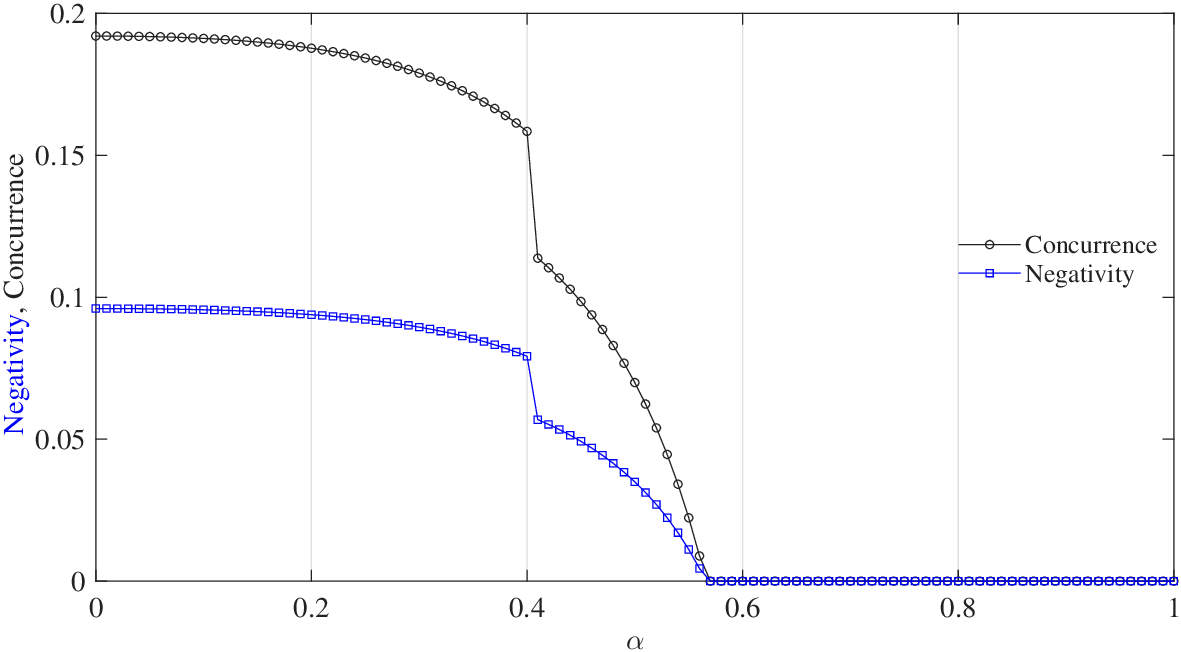}
    \caption{Reduced two-qubit concurrence and negativity of the subjacent mixture of the ground state and the first excited-state  for the two-dimensional \(J_1\!-\!J_2\) Heisenberg model on a \(4\times 4\) square lattice. Both quantities exhibit a clear discontinuity near the first quantum phase transition at \(\alpha \approx 0.4\). However, they vanish for \(\alpha \gtrsim 0.57\), and hence do not provide any signature of the second transition near \(\alpha \approx 0.6\).}
    \label{2D_con_ngtvy}
\end{figure}

From Fig.~\ref{2D_con_ngtvy}, we see that both concurrence and negativity become zero for \(\alpha \gtrsim 0.57\). Therefore, although bipartite entanglement detects the first quantum phase transition, it fails to capture the second one near \(\alpha \approx 0.6\).

{\ab Note} that {\ab the reduced two-qubit purity-corrected SRE of the ground state} of the two-dimensional \(J_1\!-\!J_2\) Heisenberg model exhibits a signature of the transition from the intermediate phase to the collinear N\'eel phase near \(\alpha \approx 0.6230\). {\gb{As shown in Fig.~\ref{2D_ground}, the reduced two-qubit purity-corrected SRE of the ground state exhibits a distinct change in curvature in the vicinity of the second quantum phase transition. The curvature changes from convex to concave, indicating a qualitative change in the evolution of the local nonstabilizer resource. Within the intermediate phase, the reduced purity-corrected SRE decreases rapidly as the frustration parameter is increased. As the system approaches the transition to the collinear N\'eel phase, however, this decreasing trend slows down, giving rise to the observed curvature change. The reduced purity-corrected SRE continues to decrease inside the collinear N\'eel phase, but with a significantly {\ab lesser} rate of change.}}

\begin{figure}
    \centering
    \includegraphics[width=\linewidth]{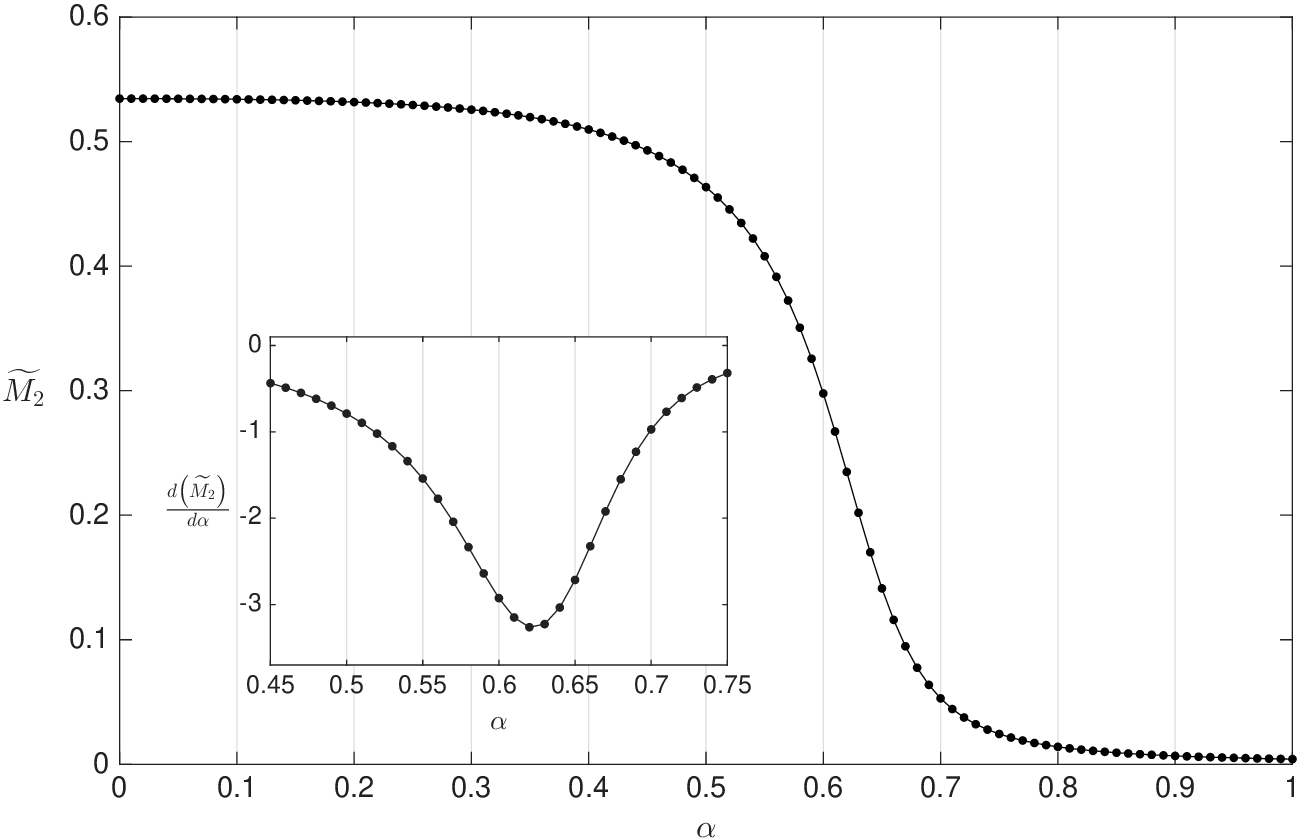}
    \caption{Reduced two-qubit {\gb{purity-corrected}} stabilizer R\'enyi entropy \(\widetilde{M_2}\) of the ground state for the two-dimensional \(J_1\!-\!J_2\) Heisenberg model on a \(4\times 4\) square lattice (\(N=16\)) with periodic boundary conditions. A transition, near \(\alpha \approx 0.6230\), is identified through a change in curvature of \(\widetilde{M_2}\). The inset shows the derivative \(d(\widetilde{M_2})/d\alpha\), whose minimum locates the transition.}
    \label{2D_ground}
\end{figure}

{\ab We note} from Fig.~\ref{ground_ent}(b) {\ab that} the reduced two-qubit bipartite entanglement in the ground state does not show any clear signature of either of the two quantum phase transitions. Therefore, in the two-dimensional \(J_1\!-\!J_2\) Heisenberg model, reduced two-qubit purity-corrected SRE provides information beyond standard bipartite entanglement {\ab indicating} the second quantum phase transition for the ground state {\ab as well as the subjacent mixed state}.
{{\gb{Although the microscopic origin of this behavior remains to be understood, the observed change in curvature suggests that the local purity-corrected SRE is redistributed as the system evolves from the intermediate frustrated phase toward the collinear N\'eel phase. Consequently, the reduced two-qubit purity-corrected SRE {of the ground state} provides a clear local signature of the second quantum phase transition, even when conventional reduced-state bipartite entanglement measures fail to do so.}}

\section{Conclusion}
\label{sec:conclusion}

In this work, we have shown that the second-order purity-corrected stabilizer R\'enyi entropy (SRE) of reduced two-qubit density matrices provides an efficient local probe of quantum phase transitions in frustrated \(J_1\!-\!J_2\) spin systems. Instead of evaluating the purity-corrected SRE of the full many-body state, whose computational cost grows exponentially with system size, we considered reduced states obtained from both the ground state and a low-temperature subjacent mixed state, defined as a statistical mixture of the ground state and the first excited state. This reduced-state approach greatly reduces the computational complexity while preserving the essential signatures of the underlying many-body quantum criticality.

{\gb{For the one-dimensional isotropic \(J_1\!-\!J_2\) Heisenberg model, we find that the reduced two-qubit purity-corrected stabilizer R\'enyi entropy (SRE) of the subjacent mixed state exhibits a clear discontinuity at the spin-fluid--dimer quantum phase transition. A finite-size scaling analysis of the discontinuity yields the thermodynamic-limit estimate \(\alpha_c(\infty)=0.24116\), in excellent agreement with the established critical value. These results demonstrate that the reduced-state purity-corrected SRE provides a reliable local probe of the quantum critical behavior in the frustrated spin chain.}}

{\gb{An interesting observation emerges when only the ground state is considered. Unlike previously studied quantum information measures, including concurrence, negativity, the generalized geometric measure (GGM), bipartite and multipartite nonlocality, and fidelity, which fail to reveal the transition in the ground state {\ab of finite size systems}~\cite{PhysRevA.70.052302,Chen2007,Biswas2014,Biswas_2020,Mondal_2023,Biswas_2024,Bao_2024}, the reduced two-qubit purity-corrected SRE exhibits a distinct change in curvature near the critical point. Specifically, the reduced purity-corrected SRE grows with an increasing rate throughout the spin-fluid phase, followed by a qualitative change in the growth trend near the onset of the dimerized phase: it continues to increase over a narrow parameter range with a decreasing growth rate, reaches a maximum, and subsequently decreases deeper inside the dimerized phase. Although the microscopic origin of this behavior remains {\ab a topic for further investigation}, it suggests that purity-corrected SRE captures aspects of the local reduced state that are complementary to those quantified by conventional quantum information measures. This enhanced sensitivity makes reduced-state SRE a promising diagnostic {\ab tool} of frustrated quantum criticality.}}

{\gb{For the {\ab richer} one-dimensional XXZ \(J_1\!-\!J_2\) model, we find that the reduced two-qubit purity-corrected stabilizer R\'enyi entropy of the ground state is generally unable to provide a reliable signature of the quantum phase boundaries. Although {\ab an inflection point} signature survives in the vicinity of the isotropic point \((\delta=1)\), it disappears as the anisotropy is increased. In contrast, the reduced two-qubit purity-corrected SRE of the subjacent mixed state successfully detects all phase boundaries of the model. The spin-fluid--dimer and dimer--N\`eel transitions are identified through discontinuities in the reduced SRE, while the spin-fluid--N\'eel transition is signaled by a sharp kink at the isotropic point. The extracted critical points accurately reproduce the known V-shaped phase boundary in the \((\alpha,\delta)\) plane, demonstrating that incorporating a small admixture of the first excited state significantly enhances the sensitivity of reduced-state purity-corrected SRE to quantum criticality in anisotropic frustrated spin systems.}}

{\gb{For the two-dimensional \(J_1\!-\!J_2\) Heisenberg model on a \(4\times4\) square lattice with periodic boundary conditions, the reduced two-qubit purity-corrected SRE when studied with the subjacent mixed state detects two distinct quantum phase transitions. The first transition, near \(\alpha\approx0.40781\), is signaled by a discontinuity in the purity-corrected SRE, whereas the second transition, near \(\alpha\approx0.6208\), is identified through a change in curvature, or equivalently by an extremum of its first derivative. In this model, purity-corrected SRE exhibits a clear advantage over conventional bipartite entanglement. While the concurrence and negativity of the subjacent mixed state detect only the first transition and vanish for \(\alpha\gtrsim0.57\), the reduced two-qubit purity-corrected SRE {\ab of the ground state as well as the subjacent state} continues to capture the second transition. Unlike reduced two-qubit bipartite entanglement {\ab of the ground state}, which fails to reveal either transition, the reduced two-qubit purity-corrected SRE of the ground state successfully identifies the second transition through a distinct change in curvature near \(\alpha\approx0.6230\). These results demonstrate that reduced-state purity-corrected SRE contains information about the underlying quantum criticality that is inaccessible to conventional bipartite entanglement measures, establishing it as a more sensitive local probe of frustrated quantum phase transitions in two dimensions.}}

{\gb{Overall, our results establish the reduced two-qubit purity-corrected stabilizer R\'enyi entropy as an effective and computationally efficient local probe of quantum criticality in frustrated spin systems. In the one-dimensional \(J_1\!-\!J_2\) model, it performs comparably to reduced two-qubit bipartite entanglement for the subjacent mixed state, while exhibiting a clear advantage in the ground state by revealing the spin-fluid--dimer transition through a change in curvature. In the two-dimensional \(J_1\!-\!J_2\) model, its advantage is even more pronounced: the reduced-state purity-corrected SRE successfully identifies the second quantum phase transition for both the subjacent mixed state and the ground state, where conventional bipartite entanglement fails. These results demonstrate that reduced-state purity-corrected SRE provides complementary information about many-body quantum criticality and can reveal phase-transition signatures inaccessible to conventional entanglement measures. A natural direction for future work is to extend this analysis to other frustrated spin systems, disordered and higher-dimensional models, and tensor-network-based simulations, as well as to compare reduced-state purity-corrected SRE with other quantum correlation measures across quantum phase boundaries.}}

\section{Acknowledgments}
G.~B. acknowledges support from the National Science and Technology Council, Taiwan, under Grant No.~114-2811-M-032-006. S.~S. acknowledges support from the Council of Scientific and Industrial Research through Grant No.~09/1336(11432)/2021-EMR-I. J.~Y.~W. acknowledges support from the National Science and Technology Council (NSTC), Taiwan, under Grant Nos.~112-2112-M-032-008-MY3, 114-2119-M-008-008,115-2119-M-008 -001, and 115-2811-M-032-005.

\section*{References}
\bibliographystyle{iopart-num}
\bibliography{PhD_references}

\end{document}